\documentclass[12pt]{article} 
\usepackage{graphicx}
\usepackage{caption}
\usepackage{mathrsfs}
\usepackage{float}
\usepackage{subcaption}
\usepackage[utf8]{inputenc}
\usepackage{longtable}
\usepackage{float}
\usepackage{booktabs}
\usepackage[english]{babel}
\usepackage[dvipsnames]{xcolor}
\usepackage{booktabs}
\usepackage{multirow}
\usepackage{ragged2e}
\usepackage{hyperref}
\definecolor{darkblue}{RGB}{0,0,120}
\hypersetup{
      colorlinks=true,
      linkcolor=darkblue,
      filecolor=darkblue,
      citecolor=darkblue,      
      urlcolor=darkblue,
      }

\usepackage[justification=centering]{caption}

\usepackage[toc,page]{appendix}
\usepackage{adjustbox}
\usepackage{multirow}
\usepackage{threeparttable, tablefootnote,booktabs, stackengine}
\usepackage{array}
\usepackage{caption}
\usepackage{placeins}

\usepackage{indentfirst}
\usepackage[margin=1in]{geometry}
\usepackage{subcaption}

\usepackage{amsmath,amsthm,amssymb,scrextend}
\usepackage{cleveref}
\usepackage{bbm}
\usepackage{mathrsfs}
\usepackage{blindtext}

\usepackage{pdflscape}
\usepackage{longtable}
\usepackage{vcell}
\usepackage{booktabs}
\usepackage{natbib}
 \bibpunct[, ]{(}{)}{;}{a}{}{,}%

\DeclareCaptionLabelFormat{andtable}{#1~#2  \&  \tablename~\upthetable}

\usepackage{setspace}
\usepackage{lipsum}
\usepackage{footmisc}

 \usepackage{spverbatim}
\usepackage{float}
\usepackage{graphicx}
\usepackage{titlesec}

\newcolumntype{x}[1]{%
>{\centering\hspace{0pt}}p{#1}}%

\DeclareMathOperator*{\argmax}{argmax}
\DeclareMathOperator*{\argmin}{argmin}

\usepackage{dsfont}
\usepackage{upgreek}

\usepackage[linesnumbered,ruled,vlined]{algorithm2e}
\SetKwComment{tcp}{\tiny // }{}

\begin{document}
\thispagestyle{empty}

\begin{titlepage}

\singlespacing

\title{Reinforcement Learning Based Computationally Efficient Conditional Choice Simulation Estimation of Dynamic Discrete Choice Models\thanks{The authors are listed in alphabetical order. We thank John Rust for insightful comments during the second author's dissertation examination. We also thank Dominique Lauga, Shasha Lu, Jaideep Prabhu, and other participants of the Marketing Seminar at the Judge Business School for their comments. We gratefully acknowledge financial support from The Turing Institute Enrichment Community Award, The Tony Cowling Foundation Award, and The Keynes Fund. All the errors are ours alone.}}
\author{Ahmed Khwaja\thanks{Judge Business School, University of Cambridge. Email: \texttt{a.khwaja@jbs.cam.ac.uk}.}
\and Sonal Srivastava\thanks{Judge Business School, University of Cambridge. Email: \texttt{ss2450@cam.ac.uk}.}
}
\date{\MakeUppercase{work in progress}\\
\vspace*{0.25cm}
\today}

\maketitle
\thispagestyle{empty}

\begin{abstract}
			\singlespacing
\noindent Dynamic discrete choice (DDC) models have found widespread application in marketing. However, estimating these becomes challenging in “big data” settings with high-dimensional state–action spaces. To address this challenge, this paper develops a Reinforcement Learning (RL)–based two-step (``computationally light'') Conditional Choice Simulation (CCS) estimation approach that combines the scalability of machine learning with the transparency, explainability, and interpretability of structural models, that is particularly valuable for counterfactual policy analysis. The method is premised on three insights: (1) the CCS (“forward simulation”) approach is a special case of RL algorithms, (2) starting from an initial state-action pair CCS updates the corresponding value function only after each simulation path has terminated, whereas RL algorithms may update for all the state-action pairs visited along a simulated path, and (3) RL focuses on inferring an agent’s optimal policy with known reward functions, whereas DDC models focus on estimating the reward functions presupposing optimal policies. The procedure’s computational efficiency over CCS estimation is demonstrated using Monte Carlos with a canonical machine replacement and a consumer food purchase model. Framing CCS estimation of DDC models as a RL problem increases their applicability and scalability to high-dimensional marketing problems while retaining both interpretability and tractability.		
			
\end{abstract}
	
\vspace{0mm}\noindent
		\textbf{Keywords:} \\ Dynamic Discrete Choice, Reinforcement Learning, Markov Decision Process, Conditional Choice Simulation, Two-step Estimation, Forward Simulation

\end{titlepage}

\clearpage 
\doublespacing
\pagenumbering{arabic} 
	\newpage

\section{Introduction}
Since their introduction to marketing research with pioneering applications that include consumer learning about brand quality \citep{erdem1996decision}, response to promotions \citep{gonul1996estimating}, and catalog mailing decisions \citep{gonul1998optimal}, dynamic discrete choice (DDC) models in a Markov Decision Process (MDP) framework have found widespread application in studying fundamental marketing problems.\footnote{See for example, work related to optimal intertemporal pricing strategy in durable goods markets \citep{nair2007intertemporal}, demand in reward programs \citep{hartmann2008frequency}, household behavior when brand attributes are uncertain but price and advertising signal quality \citep{erdem2008dynamic}, replacement and purchase decisions in technology industry \citep{gordon2009dynamic,sriram2010investigating}, observational learning in US kidney market \citep{zhang2010sound}, sales-force compensation strategies \citep{misra2011structural,chung2014bonuses,kim2022structural}, endogenizing market structure and long-run innovation to understand the relationship between competition and innovation \citep{GoettlerGordon2011,GoettlerGordon2014}, framing service channel allocation as a dynamic programming problem based on customer preferences \citep{SunLi2011}, firm's cross-selling strategies to maximize long-term profit \citep{LiSunMontgomery2011}, consumer behavior in seasonal goods markets \citep{soysal2012demand}, decision-making between products and bundles \citep{derdenger2013dynamic}, seller reputations in freelance sites \citep{yoganarasimhan2013value}, adoption and usage decisions of ATM cards \citep{yang2014dynamics}, ability of consumers to optimally allocate usage under multipart tariffs \citep{gopalakrishnan2015consumer}, dynamic model of addiction to cigarettes to evaluate how consumers respond to price changes \citep{GordonSun2015}, thirst management \citep{huang2015short}, evolution of brand preferences of new consumers \citep{che2015consumer}, effectiveness of green technology adoption policies \citep{bollinger2015green}, scheduling detailing activities in pharmaceutical markets \citep{liu2016empirical}, insurance plan decisions \citep{mehta2017dynamic}, effectiveness of reward programs in travel industry \citep{rossi2018lower}, monitoring costs and consumer dissatisfaction associated with overdraft fees \citep{liu2018analyzing}, dynamic oligopoly pricing model in the presence of switching costs \citep{cosguner2018dynamic}, adverse selection in car insurance markets \citep{jeziorski2019skimming}, intertemporal price discrimination in e-readers and e-books \citep{li2019intertemporal}, and owner decisions in peer-to-peer sharing marketplaces \citep{yao2023dynamic}.} DDC models are well-suited to studying many marketing problems because these are inherently dynamic as the decisions are sequential and depend on past experiences, present conditions and expectations about future consequences. For example, consumer choices about brand adoption and loyalty, or switching providers are based not only on previous experience, and current benefits and costs but also on expectations about these in the future. Firms, in turn, design pricing, promotion, and retention strategies with the explicit goal of shaping these dynamic decisions. Accounting for such dynamic aspects of consumer and firm decisions is therefore essential for understanding long-term customer value creation and for evaluating the effectiveness of various marketing strategies.

While DDC models provide insights about decision-theoretic primitives and counter-factual policy analysis \citep{lucas1976econometric},\footnote{Seminal applications in economics include research related to decisions about employee retention \citep{gotz1984dynamic}, occupational matching \citep{miller1984job}, fertility \citep{wolpin1984estimable,montgomery1988dynamic}, patent renewal \citep{pakes1986patents}, engine replacement \citep{rust1987optimal} and job search, employment and retirement \citep{wolpin1987estimating,gonul1989dynamic,eckstein1989dynamic,berkovec1991job}. For excellent reviews and critiques of applications in economics see e.g.,~\cite{eckstein1989specification}, \cite{wolpin1996public}, \cite{blundell2006advances}, \cite{aguirregabiria2010dynamic}, \cite{keane2010perspective}, \cite{keane2010structural}, \cite{rust2010comments}, \cite{keane2011structural}, \cite{arcidiacono2011practical}, \cite{wolpin2013limits}, \cite{low2017use} and \cite{blundell2017have}.}  their estimation becomes prohibitively costly and infeasible, limiting their application to problems of practical interest as the state space and number of choices grow. This is particularly true with the availability of bigger and more granular data sets \citep{naik2008challenges, chintagunta2016marketing}. This is because traditional DDC model estimation approaches, like the Nested Fixed Point (``NFXP'') algorithm \citep{rust1987optimal}, solve the full dynamic programming (DP) problem to compute value functions accurately but suffer from the ``curse of dimensionality.''
The ``computationally lighter'' two-step ``forward simulation'' Conditional Choice Simulation (CCS) estimator \citep{hotz1994simulation}, and its companion Conditional Choice Probability (CCP) estimator \citep{hotz1993conditional}, eliminate the demanding step of repeated state-action space sweeps for fixed point computation as in NXFP by exploiting the mapping between normalized value functions and CCPs.\footnote{For an excellent review of CCP-based estimation of DDC models see e.g., \cite{arcidiacono2011practical}.} However, both these, and the CCS especially, rely on sufficiently long simulation paths for accurate computation of value functions that may not be most computationally efficient. Although simplifying such problems can improve scalability, it often comes at the cost of accuracy and rigor, both of which are essential for generating reliable insights \citep{naik2008challenges, chintagunta2016marketing}.

We propose algorithms that integrate Reinforcement Learning (RL) with two-step CCS estimation to improve computational efficiency without compromising accuracy. RL has emerged as one of the most powerful tools in Artificial Intelligence (AI) for automating decision-making and control. It has been successfully applied to a wide range of complex tasks, from self-learning systems that master games like Go \citep{silver2018general} to fine-tuning Large Language Models to better align with human preferences, factual accuracy, and safety \citep{christiano2017deep, ouyang2022training}. Like DDC models, RL addresses sequential decision-making problems using the MDP framework. However, while RL assumes the reward function is known and focuses on learning optimal policies, the DDC approach estimates reward functions (model primitives) from observed behavior, assuming optimal policies.

Our algorithms rely on three insights: (1) we first show that the CCS (``forward simulation'') approach is a special case of RL “Monte Carlo” and “Temporal Difference” algorithms, (2) starting from an initial state-action pair CCS updates the corresponding value function \textit{only after} each simulation path has terminated, whereas RL methods may update the value functions for \textit{all} the state-action pairs that are visited along a simulated path, and (3) although the RL and DDC approaches have different inferential goals, \textit{both} focus on accurate computation of value functions. We develop two versions: (1) RLMC-CCS based on RL's ``Monte Carlo'' (MC) \citep{singh1996reinforcement}, and (2) RLTD-CCS based on Temporal Difference (TD) \citep{sutton1988learning} algorithms.

The main intuition behind our methods is that within each forward simulated path, a sequence of states and actions are visited, often multiple times. By treating these visits as the initial nodes (state-action pair) for computing the value functions as well, many more updates can be made in each simulated path. As the number of forward simulation paths increase, the number of different state-action visits also increases. The first algorithm, RLMC-CCS, treats each full path as a set of multiple sub-paths, each starting from individual state-action pairs visited along the path. However, when simulated paths are kept short to reduce the computational effort, terminating after only a few time steps, the estimates are shown to be less accurate. The second, more promising, RLTD-CCS algorithm provides highly accurate estimates even for shorter paths. Just like CCS, RLTD-CCS computes value functions over forward simulated paths. However, instead of relying on the total discounted returns from the entire length of a path, RLTD-CCS looks one or more steps ahead and uses the most recent value function update of the next state-action pair (or of the state-action pair encountered after simulating a few steps ahead). This converts the value function computation into an iterative update process, where value function update for a state-action pair is related (or, ``bootstrapped" in RL jargon) to the latest value function of the future state-action pair(s), which is similar to fixed-point iterations in NFXP. The large number of bootstrapped updates help to converge to accurate estimates of value functions. Within the RLTD-CCS algorithm, by using a value function ``learning'' parameter, the value function updates get naturally linked to the general class of stochastic machine learning algorithms (\cite{robbins1951stochastic}, \cite{kiefer1952stochastic}).

Using Monte Carlo studies, we assess our approach in two settings: (i) a canonical machine replacement model \citep{rust1987optimal}, and (ii) a  consumer food choice model \citep{huang2015short}. For both the models, we use the root-mean-square error (RMSE) between the modeled and estimated structural parameters as a metric to compare the estimation accuracy between the different algorithms. The RLTD-CCS outperforms both RLMC-CCS and CCS, achieving up to 3-4 orders of magnitude greater estimation accuracy using simulation paths that are upto ten times shorter than CCS. The performance remains robust even for large state-action spaces (about 5.77 million states), with RLTD-CCS surpassing CCS by a speed factor of 6 to 14 times. We also demonstrate that RLTD-CCS is less sensitive to discount ($\beta$), achieving a 70 factor speed advantage over CCS in estimation accuracy when the discount factor approaches one.

Our work contributes to the literature that is focused on estimating single agent DDC models in an MDP framework (although, it may also indirectly contribute to the literature on multi-agent models such as games of interaction between strategic agents).\footnote{For example, \cite{aguirregabiria2007sequential}, \cite{bajari2007estimating}, \cite{pakes2007simple} and \cite{pesendorfer2008asymptotic} build on the CCP and CCS formulations to introduce methods for estimating dynamic games of strategic interactions. See e.g., \cite{aguirregabiria2013recent} and \cite{aguirregabiria2021dynamic} for comprehensive reviews of the literature on estimating dynamic games.}  There has been tremendous progress in this field over the last several decades since the work of \cite{rust1987optimal} introduced the NXFP algorithm, and several approaches have been suggested to overcome this computational challenge. In addition to the CCP \citep{hotz1993conditional} and CCS \citep{hotz1994simulation} methods, these include computing value functions employing the Gittins Index \citep{miller1984job,eckstein1988empirical}, using Monte Carlo integration over a sub-set of states and then interpolating using a regression function \citep{keane1994solution}, using random Bellman Operators on a sub-set of grid-points \citep{rust1997using}, iterating over policy functions instead of value functions \citep{aguirregabiria2002swapping}, use of Bayesian Markov Chain Monte Carlo based algorithms \citep{imai2009bayesian, norets2009inference}, relying on the properties of finite dependence or renewal \citep{arcidiacono2011conditional}, using (equilibrium) constrained optimization \citep{su2012constrained}, approximation using artificial neural networks \citep{norets2012estimation}, using the Endogenous Grid-point Method (EGM) \citep{iskhakov2017endogenous}, and using data driven state-space partitioning \citep{barzegary2022recursive}. Most often, these methods provide an exact solution only for a sub-set of the state-space and assign an approximate value to the remaining (excluded) states. However, it is unclear whether these methods are generally applicable when the state and action space becomes large and the models more complex in being high-dimensional \citep{britton2021revisiting}. In contrast, our approach is different as we do not approximate the value functions by using a subset of either the value functions or the state-action space, but rather use more computationally efficient simulation-based algorithms that like CCS operate on the entire observed state-action space. Our work extends the CCS based estimation methods by proposing new two-step forward simulation estimators. Another advantage of our approach is that it can be combined with any existing estimation method which is appropriate for CCP or CCS based estimation.\footnote{For example, our algorithms work well within the EM algorithm extension introduced by \cite{arcidiacono2011conditional} for estimating DDC models with unobserved heterogeneity.}

Further, we contribute to the growing literature that uses machine-learning (ML) to estimate structural parameters (see e.g., \cite{iskhakov2020machine}). These include using a non-parametric causal forest based approach to estimating heterogeneous treatment effects \citep{wager2018estimation}, decision-trees for model selection \citep{schwartz2014model}, and neural-networks for estimating (static) discrete choice models \citep{wei2025estimating}. Our approach bridges the gap between the DDC and RL literatures and leverages on the similarities in value function computation in the two domains. As we focus on integrating computationally efficient machine learning (RL in our case) into DDC models, we propose fast, scalable algorithms to estimate structural models that are inherently transparent, interpretable, and explainable, rather than attempting to make machine learning models themselves more interpretable \citep{rudin2019stop}.  

Our work also contributes to the growing Inverse Reinforcement Learning (IRL) \citep{russell1998learning, ng2000algorithms} literature, where the central objective is to infer a reward function from observed expert behavior and then derive an optimal policy. IRL can be viewed as a form of Imitation or Apprenticeship Learning \citep{abbeel2004apprenticeship,ciosek2021imitation}, but differs from direct imitation in its attempt to recover the underlying preferences driving observed choices. A well-known challenge in IRL is identifiability: the problem is ill-posed, as multiple reward functions can explain the same behavior. One influential approach, Maximum Entropy IRL \citep{ziebart2008maximum}, partially addresses this be selecting the reward function that explains the behavior while maximizing entropy over the policy space. Building on this and the ``Behavioral Cloning'' \citep{torabi2018bco} approach, \cite{kang2025gradients} use ``anchor action'' with fixed reward to regularize estimation, and report a scalable algorithm that doesn't require knowledge of transition dynamics. However, the estimated reward functions in these approaches are not structural. Separately, \cite{lee2025prep} apply Adversarial IRL \citep{fu2018airl} to high-dimensional content consumption data, and use computationally intensive simulated roll-outs to closely align the predicted and observed behavior. In contrast, our work embeds RL algorithms in CCS and estimates structural DDC models in a computationally efficient way from observed behavior. This hybrid approach retains the interpretability and policy structure of DDC models while gaining the scalability of RL, offering novel and practical alternative within the IRL landscape.

The work most closely related to ours is \cite{adusumilli2019temporal}. In this work, function approximation of value functions within a linear semi-gradient algorithm is used to solve continuous state space models. This approach was adapted from the semi-gradient TD(0) algorithm (\cite{sutton2018reinforcement}). In a similar work, basis function approximation was used to minimize the TD approximation error to estimate DDC models~\citep{imaizumi2015approximation}. However, methods that use function approximation and TD learning converge to a unique ``TD fixed point,'' the asymptotic error of which is $\frac{1}{1-\beta}$ times the smallest error that one could obtain using function approximation with an unbiased MC (e.g., CCS) method \citep{sutton2018reinforcement}. As $\beta$ becomes close to 1, this error increases. Our method, on the other hand, does not approximate value functions, but computes value functions by using CCS simulations in a more computationally efficient way. We report that the estimation accuracy of our methods hold even when $\beta$ approaches 1, while getting even better speed and memory advantage than CCS. 

The remainder of the paper is organized as follows. In the next section, we introduce reinforcement learning of value functions. We then present Monte Carlo studies, and finally, we conclude with a summary of our findings.

\section{Reinforcement Learning of Value Functions} \label{sec1}
\noindent We use a canonical single agent DDC model in the MDP framework \citep{rust1987optimal} to illustrate our method that we briefly describe next.

\subsection{A Canonical DDC Model \& CCS Estimation}
\noindent An agent takes sequential decisions in periods $t=1,2,...,T \leq \infty$. In the period $t$, the agent observes a discrete state $s_t\in \mathcal{S} = \left\lbrace 1,2,...,S_t \right\rbrace$ and then chooses a discrete action $a_t \in \mathcal{A} = \left\lbrace 1,2,...,J
\right\rbrace$. All the actions are assumed to be mutually exclusive in any period. Before choosing an action, the agent is assumed to also observe a vector of states (unobservable to the researcher) which are denoted by $\epsilon_t = \left( \epsilon(a_t=1),\epsilon(a_t=2),...,\epsilon(a_t=J) \right)$.\footnote{In the absence of the unobservable (to researcher) state $\epsilon_t$, the agent's decisions will be a deterministic function of the observed (by both agent and researcher) state variable $s_t$. This implies that two agents, in the identical state $s_t$, will take same decision. However, this is seldom the case in the data. Including an unobserved random utility component $\epsilon_t$ allows the model to rationalize the data by permitting different decisions by agents in the identical state $s_t$. For example, in the engine replacement model of \cite{rust1987optimal}, the model will imply that the superintendent will take the identical decision to replace the engine (or not) for all the buses in the same state (mileage). However, the data will clearly contradict this.} 
Upon taking an action, the agent receives an immediate payoff or utility, $U(s_t,\epsilon(a_t),a_t; \theta_u)$ where $\theta_u$ are the structural parameters of the utility function that need to be estimated. The objective of the agent in any period is to maximize the present discounted value (PDV) of the future expected utility stream:
\begin{equation}
 \mathbb{E}\left[\sum_{\tau=t}^{T} \beta^{\tau-t} U\left(s_\tau, \epsilon(a_\tau),a_\tau; \theta_u\right) \bigg| s_t,\epsilon_t; \theta_F \right],
    \label{eq:expected_utility}
\end{equation}
where $\mathbb{E}$ is the expectation operator with expectations over future observable and unobservable states. The $\beta \in (0,1)$ is the discount-factor, and $(\theta_u, \theta_F)=\theta$ are the parameters to be estimated, where $\theta_F$ are parameters related to state transitions.

Several assumptions have been made in the literature to make the dynamic discrete-choice model \citep{rust1987optimal} empirically tractable. We begin by briefly discussing these.

\noindent\textbf{Assumption 1} \textit{Markovian policy}: Only current states are relevant for current action and summarize any relevant information from past actions and states. However, this is not as restrictive an assumption as the information from the previous periods can also be accommodated in modeling the decision process by modifying the definition of the state space. As \cite{rust1994structural} quoting \cite{10.5555/26970} states, “the well-known trick of `expanding the state space' can be used to transform an $N$\textsuperscript{th} order Markov process into a 1\textsuperscript{st} order Markov process.''

\noindent\textbf{Assumption 2} \textit{Conditional Independence}: The joint transition probability can be factored as following:
\begin{equation}
	\mathit{p}(s_{t+1}, \epsilon_{t+1} \big| s_{t}, \epsilon(a_t), a_{t}; \theta_F) = \mathit{p}(\epsilon_{t+1} \big| s_{t+1}) \mathit{p}(s_{t+1} \big| s_{t}, a_{t}; \theta_F) .
    \label{eq:ci_assumption}
\end{equation}
This condition imposes restriction on the dynamics of the decision process by implying that the occurrence of $s_{t+1}$ depends only on the current state $s_t$ and action $a_t$. This condition further implies that the unobserved states are serially independent across periods and conditional only on the current state. While this is a strong restriction, it can be relaxed to allow for serial correlation.

\noindent\textbf{Assumption 3} \textit{Additive separability}: Each single-period utility is assumed to be additive separable in its deterministic and stochastic components, such that:
\begin{equation}
    U(s_t,\epsilon(a_t),a_t; \theta_u) = u(s_t,a_t; \theta_u) + \epsilon(a_t).
    \label{eq:additive_sep}
\end{equation}
To maximize the PDV of the stream of utilities, the agent follows an \textit{optimal policy} comprising a sequence of actions, $\left\lbrace a_\tau(s_\tau,\epsilon_\tau; \theta)\right\rbrace_{a_\tau\in \mathcal{A}, \tau=1,...,\infty}$, in each period.
The PDV of the expected utility stream at each state when the agent follows an optimal policy is represented by the state's value function:
\begin{equation}
	V(s_t, \epsilon_t; \theta) \doteq \max_{a_t \in \mathcal{A}} \mathbb{E}\left[\sum_{\tau=t}^{T} \beta^{\tau-t} U\left(s_\tau, \epsilon(a_\tau),a_\tau; \theta_u\right)  \big| s_t, \epsilon_t; \theta_F\right] .
    \label{eq:value_functions}
\end{equation}
The optimal decision-rule can be expressed as:
\begin{equation}
   a_t^{*}(s_t, \epsilon_t;\theta) = \argmax_{a_t \in \mathcal{A}} \mathbb{E}\left[\sum_{\tau=t}^{T} \beta^{\tau-t} U\left(s_\tau, \epsilon(a_\tau),a_\tau; \theta_u\right)  \big| s_t, \epsilon_t; \theta_F\right] .
    \label{eq:policy_equation}
\end{equation}
For notational simplicity we drop the $t$ subscript unless indicated. Following convention we use $s^{\prime},\epsilon^{\prime}$, and $a^\prime$ to indicate the states and action in the next period. The value function can be simplified into a single period representation by using the \textit{Bellman Equation} and the additive separability assumption on the single-period utility as follows: 
\begin{equation}
    V(s, \epsilon; \theta) = \max_{a \in \mathcal{A}} \left[u(s,a; \theta_u) + \epsilon(a) + \beta \mathbb{E} \left[V(s^\prime,\epsilon^\prime; \theta) \big| s,\epsilon(a), a\right] \right]
    \label{eq:bellman_eq}
\end{equation}
Further, the \textit{choice-specific value function} can be defined as:
\begin{equation}
    V(s,\epsilon(a), a; \theta) \doteq u(s,a; \theta_u) + \epsilon(a) + \beta \mathbb{E} \left[V(s^\prime,\epsilon^\prime; \theta) \big| s,\epsilon(a), a\right] .
    \label{eq:csvf_def}
\end{equation}
The value function can be written in terms of the choice-specific value functions as follows:
\begin{equation}
    V(s,\epsilon; \theta) = \max_{a \in \mathcal{A}} \left[V(s,\epsilon(a), a; \theta) \right] .
    \label{eq:vf_csvf}
\end{equation}
We can extend the additive separability assumption to the choice-specific value functions by decomposing it into deterministic and stochastic components:
\begin{equation}
    V(s,\epsilon(a), a; \theta) = v(s,a; \theta) + \epsilon(a) .
    \label{eq:csvf_determ_stochast}
\end{equation}
The decomposed choice-specific value function formulation can be used to derive the conditional choice probability (CCP), $\pi(\cdot)$, for selecting a particular action $m$ in state $s$ by integrating out the stochastic unobserved states:
\begin{equation}
    \begin{split}
        \pi(a = m | s; \theta) &= \text{Prob} \left\lbrace V(s, \epsilon(m),m; \theta) > V(s,\epsilon(j), j; \theta), \forall \; \; a = j \neq m \right\rbrace \\
        &= \int \mathbb{I}(\epsilon(j) < \epsilon(m) + v(s,m; \theta) - v(s,j; \theta), \forall \; \; a= j \neq m) dG(\epsilon | s) \; ,
    \end{split}
    \label{eq:ccp_integral}
\end{equation}
where, the $J$-dimensional integral is taken over the joint density of the vector $\epsilon$.

The computational burden of evaluating the integral, when combined with the expectation operation over the entire state-action space comprising $\left(s, \epsilon \right)$ is extreme. While defining a finite $\mathcal{S}$ relieves some of the computation challenge, integrating over the stochastic unobserved states is still resource consuming.

\noindent\textbf{Assumption 4} \textit{Error Distribution}: In order to make the model more tractable, the convention has been to assume that the unobserved states are IID across choices and periods and given by the Type-1 Multivariate Extreme Value (MEV) distribution \citep{mcfadden1973conditional,rust1987optimal}.\footnote{We follow the long literature starting with \cite{mcfadden1973conditional} in using the MEV distribution as do \cite{hotz1993conditional}, \cite{hotz1994simulation} and \cite{rust1987optimal}. See e.g., \cite{keane1997career} for a model that uses the normal distribution for the unobserved states.} This assumption, when combined with the conditional independence assumption, reduces the CCP into a familiar logit-style expression:
\begin{equation}
    \pi(a = m | s; \theta) = \frac{e^{v(s,m; \theta)}}{\sum_{j \in \mathcal{A}} e^{v(s,j; \theta)}} .
    \label{eq:ccp_mev}
\end{equation}
However, unlike the static case where the logit expression only includes the current single-period utility, the ``dynamic logit''  expression includes the deterministic component of the choice specific value function (Equation~\ref{eq:csvf_determ_stochast}) that depends on the PDV of the expected future utilities thereby making CCPs more complicated to compute.

CCS is predicated on the insight that instead of solving the dynamic programming problem to compute the value functions as the fixed point of a Bellman equation (NFXP) these may be computed directly using a ``forward simulation'' approach \citep{hotz1994simulation}. 

The ``two-step" CCS estimation procedure involves a first step of obtaining non-parametric estimates of both the transition probability, $\hat{\mathit{p}}(s_{t+1} \vert s_t, a_t; \hat{\theta}_F)$, and CCPs, $\hat{\pi}(a_t \vert s_t)$, directly from the data. In the second step, for a given guess $\hat{\theta}_u$, the corresponding value functions are computed using ``forward simulations'' (based on  $\hat{\mathit{p}}(s_{t+1} \vert s_t, a_t; \hat{\theta}_F)$ and $\hat{\pi}(a_t \vert s_t)$) that are in turn used to compute ``simulated (or predicted)'' CCPs, $\tilde{\pi}(a_t \vert s_t; \hat{\theta}_u)$.  

A Minimum Distance Estimator (MDE) may then be used for inferring parameters that minimize the distance between the ``simulated'' and ``directly estimated'' choice probabilities (alternatively, for value functions):
\begin{equation}
    \hat{\theta}_u^{i+1} = \argmin_{\theta_u} || \tilde{\pi}(a_t \vert s_t; \hat{\theta}_u^i, \hat{\theta}_F) - \hat{\pi}(a_t \vert s_t) || \; ,
\end{equation}
where, $i$ denotes the iteration index and $\hat{\theta}_F$ represents first stage estimates. We follow \cite{bajari2007estimating} and a related stream of work that uses an MDE estimator. In their study, \cite{hotz1994simulation} used a Method of Simulated Moments estimator \citep{McFadden1989}.

Each forward simulated path consists of a sequence of state-action pairs and, in the case of infinite time-horizon problems, has a length of $T_{\text{end}}$ beyond which, due to discounting and averaging, the simulation errors are insignificant (see e.g., \cite{bajari2007estimating}, footnote 9, p.1343.). For each (forward) simulated path $k$ starting from an initial state-action pair $(s,a)$, the \textit{path-specific} deterministic component of the choice-specific value function can then be computed after simulating the full path using the following expression:
\begin{multline}
    \tilde{v}^k(s,a;\hat{\theta}_u^i, \hat{\theta}_F) = u(s,a;\hat{\theta}_u^i) + \beta \bigg[u(s^\prime, a^\prime; \hat{\theta}_u^i) + \gamma - \log{(\hat{\pi}(a^\prime | s^\prime))} + \\
    \beta \Big[ u(s^{\prime \prime}, a^{\prime \prime}; \hat{\theta}_u^i) + \gamma - \log{(\hat{\pi}(a^{\prime\prime} | s^{\prime\prime}))} + \\
    \beta \big[...+\beta [u(s_{T_{\text{end}}},a_{T_{\text{end}}}; \hat{\theta}_u^i) + \gamma - \log{(\hat{\pi}(a_{T_{\text{end}}} | s_{T_{\text{end}}}))}]...\big]\Big]\bigg]\;,
    \label{eq:ccs_gen_notation}
\end{multline}
where, the notation $^\prime, ^{\prime\prime}, ...$ represents the sequence of forward simulation-steps using $\hat{\mathit{p}}(\cdot; \hat{\theta}_F)$ and $\hat{\pi}(\cdot)$ following the initial $(s,a)$. The average value after computing the returns from $K$ paths can then be written as:
\begin{equation}
    \bar{v}^K(s,a; \hat{\theta}_u^i, \hat{\theta}_F) = \frac{1}{K}\sum_{k=1}^K \tilde{v}^k(s,a; \hat{\theta}_u^i, \hat{\theta}_F)
\end{equation}

This CCS approach is ``computationally lighter'' than the NFXP as it skips the fixed-point iterations. A complementary discussion on value function computation and estimation steps in NFXP and CCS is provided in the online appendix. The main computational burden in the CCS approach arises from the forward simulations and the subsequent averaging operation. The pseudocodes for CCS, and our proposed RLMC-CCS and RLTD-CCS algorithms are provided in the online appendix.

\subsection{CCS \& Step-wise Learning}
\noindent \textbf{First Insight:} The first insight is that CCS ``forward simulation'' approach is a special case of RL ``Monte Carlo'' and ``Temporal Difference'' algorithms.

\noindent \textbf{Second Insight:} A second insight is that RL ``Monte Carlo'' and ``Temporal Difference'' algorithms compute the value functions for not only the initial state-action pair but also for the subsequently visited state-action pairs along the path.

To demonstrate this and also how choice-specific value functions can be ``learned'' and updated efficiently, we begin by showing how CCS can be re-interpreted in the RL framework.\footnote{A brief discussion on RL methods and their comparison with DDC modeling is provided in the online appendix.}

In CCS, $k=1,\ldots,K$ paths are simulated starting from $(s,a)$ and for each path a value function is computed and then averaged over the $K$ stored values. We can break the process of computing the overall mean into individual ``update'' steps:
\begin{equation}
    \begin{split}
        \underbrace{\bar{v}(s,a; \hat{\theta}_u^i, \hat{\theta}_F)[\Lambda+1]}_{\text{New value}} &= \frac{1}{K}\sum_{k=1}^K \tilde{v}^k(s,a; \hat{\theta}_u^i, \hat{\theta}_F) \\
        &= \frac{1}{K}\left(\tilde{v}^K(s,a; \hat{\theta}_u^i, \hat{\theta}_F)[\Lambda+1] + \frac{K-1}{K-1} \sum_{k=1}^{K-1} \tilde{v}^k(s,a; \hat{\theta}_u^i, \hat{\theta}_F) \right) \\
        &= \underbrace{\bar{v}(s,a; \hat{\theta}_u^i, \hat{\theta}_F)[\Lambda]}_{\text{Old value}} + \frac{1}{K}\underbrace{\left(\tilde{v}^K(s,a; \hat{\theta}_u^i, \hat{\theta}_F)[\Lambda+1] - \bar{v}(s,a; \hat{\theta}_u^i, \hat{\theta}_F)[\Lambda] \right)}_{\text{Update error}}
    \end{split}
    \label{eq:ccs_step_wise}
\end{equation}
where, $[\Lambda+1], [\Lambda], ...$ are the global computer times when the value from a new forward simulated path is calculated.
Equivalently this entails weighting the difference between the most recent (at $[\Lambda+1]$) path-specific value ($\tilde{v}^K(s,a; \hat{\theta}_u^i, \hat{\theta}_F)[\Lambda+1]$) and the ``old'' (at $[\Lambda]$) mean value ($\bar{v}(s,a; \hat{\theta}_u^i, \hat{\theta}_F)[\Lambda]$), and then adding this ``error'' to the ``old'' mean value. Hence, we \textit{learn} the ``new'' mean value as a forward simulated path value becomes available. This is more memory efficient as instead of storing $K$ values to compute the mean, only $2$ memory units are needed, one to maintain the mean value and the other to store the latest simulated path value. Thus, CCS can be re-interpreted in a step-wise learning formulation that is the underlying foundation of several RL algorithms.

The weight $1/K$ ensures that as $K \rightarrow \infty$, asymptotically the impact of the update error on the mean value becomes negligible. The step-wise updates can be formalized as the key learning equation in RL:
\begin{equation}
    \text{New value} = \text{Old value} + \alpha(k) \text{ }(\text{Update Error})\;,
    \label{eq:learning_equation}
\end{equation}
where, $\alpha(k)$ is called the \textit{learning parameter} or the \textit{step-size parameter}. For example, in Equation~\ref{eq:ccs_step_wise}, $\alpha(k) = 1/k$. This is also the fundamental building block of various machine learning methods like the stochastic gradient descent algorithm. The rationale for the learning equation, and the value of the learning parameter, comes from the convergence properties of iterative stochastic approximation processes that were first proposed by \cite{robbins1951stochastic} and later extended by \cite{dvoretzky1956stochastic}. Provided the learning parameter has the following two properties: (1) $\sum_k \alpha(k) = \infty$, and (2) $\sum_k \alpha^2(k) < \infty$, then the iterative update procedure converges with probability 1. Instead of a varying learning rate, we can have a small constant value ($\alpha \in (0,1)$) that ensures that all the update errors are given the same weight. While a dynamic learning parameter that diminishes with each time-step satisfies both the conditions, a constant parameter only strictly satisfies the first condition. This might result in a noisy convergence with the mean getting affected by every new update. However, a constant learning parameter has been shown to accelerate the rate of convergence. Using a constant learning parameter is particularly useful for non-stationary MDPs, as commonly assumed in RL applications \citep{sutton2018reinforcement}. 

\subsection{RL Monte Carlo Methods \& RLMC-CCS}
\noindent To provide more intuition that CCS is a special case of RL Monte Carlo, consider an infinite-horizon MDP with two actions, $a_t = 0,1$, and finite states $s_t=1,\ldots,S$. A sample forward path starting from the initial state-action pair $(1,0)$ could be:
\begin{equation*}
    k^{\text{th}} \text{ path: }( 1,0 )^{\dagger}_{1} \rightarrow (2,0)_{2} \rightarrow (3,1)_{3} \rightarrow (1,0)_{4} \cdots \rightarrow (3,0)_{T_{\text{end}}-1} \rightarrow (4,0)_{T_{\text{end}}}  
\end{equation*}
Here, the symbol $\dagger$ in the superscript represents the initial state-action pair while the numbers in the subscript indicate the simulation step. Using CCS the path-specific choice-specific value function can be computed using Equation~\ref{eq:ccs_gen_notation} as:
\begin{equation}
    \tilde{v}^k(1,0; \hat{\theta}_u^i, \hat{\theta}_F) = u(1,0; \hat{\theta}_u^i) + \beta \left[ u(2,0; \hat{\theta}_u^i) + \gamma - \log{\left( \hat{\pi}(0 \big| 2) \right)} + \beta \left[ \cdots \right]\right]
\end{equation}
\noindent \textbf{Second Insight (RLMC):} This leads to our second insight, that our RLMC-CCS algorithm, premised on the RL Every-Visit MC algorithm, computes the value function for not only the initial state-action pair $(1,0)$ but also value functions for all subsequently visited state-action pairs along the path, $(2,0),(3,1),\ldots$. In RLMC-CCS, a forward path can be thought to be made up of several sub-paths (\textit{i.e.}, a total of $T_{\text{end}}$ sub-paths), each originating at different simulation-steps, $t=1,2,\ldots,T_{\text{end}}$. The discounted returns corresponding to  each sub-path can be used to compute value functions. If a state-action pair is visited multiple times in a path, then the sub-path specific value functions from each of the sub-paths can be used to update the mean value multiple times. To illustrate with the sample path example above, it is seen that the path can be further sub-divided into possible sub-paths:
\begin{equation*}
    \begin{split}
        & k^{\text{th}} \text{ path, }1^{\text{st}} \text{ sub-path: } ( 1,0 )^{\dagger}_{1} \rightarrow (2,0)_{2} \rightarrow (3,1)_{3} \rightarrow (1,0)_{4} \cdots \rightarrow (3,0)_{T_{\text{end}}-1} \rightarrow (4,0)_{T_{\text{end}}}\\
        & k^{\text{th}} \text{ path, }2^{\text{nd}} \text{ sub-path: } (2,0)^{\dagger}_{2} \rightarrow (3,1)_{3} \rightarrow (1,0)_{4} \cdots \rightarrow (3,0)_{T_{\text{end}}-1} \rightarrow (4,0)_{T_{\text{end}}}\\
        & \vdots\\
    \end{split}
\end{equation*}
For each sub-path, an equivalent choice-specific value function can be computed:
\begin{equation}
    \begin{split}
    &\tilde{v}^{k,1}(1,0; \hat{\theta}_u^i, \hat{\theta}_F) = u(1,0; \hat{\theta}_u^i) + \beta \left[ u(2,0; \hat{\theta}_u^i) + \gamma - \log{ (\hat{\pi} \left(0 \big|2\right))} + \beta \left[\cdots \right]\right] \\
    &\tilde{v}^{k,2}(2,0; \hat{\theta}_u^i, \hat{\theta}_F) = u(2,0; \hat{\theta}_u^i) + \beta \left[ u(3,1; \hat{\theta}_u^i) + \gamma - \log{ (\hat{\pi} \left(1 \big|3 \right))} + \beta \left[\cdots \right]\right] \\
    & \vdots \\
    \end{split}
\end{equation}

The RLMC-CCS algorithms updates can be formally written as:
\begin{equation}
    \tilde{v}(s,a; \hat{\theta}_u^i,\hat{\theta}_F)[\Lambda+1] = \tilde{v}(s,a; \hat{\theta}_u^i, \hat{\theta}_F)[\Lambda] + \alpha \left(\tilde{v}^{k,l}(s,a; \hat{\theta}_u^i, \hat{\theta}_F)[\Lambda+1] - \tilde{v}(s,a; \hat{\theta}_u^i,\hat{\theta}_F)[\Lambda] \right) \;,
    \label{eq:RLMC_general_form}
\end{equation}
where, $\tilde{v}^{k,l}(s,a; \hat{\theta}_u^i,\hat{\theta}_F)$ is computed for the $l$\textsuperscript{th} sub-path (with its first state-action pair as $(s,a)$) derived from the $k$\textsuperscript{th} simulated path.
The learning parameter $\alpha$ is $\frac{1}{\kappa+1}$ where $\kappa$ is the number of times $(s,a)$ has been visited across all the previous paths.

\noindent \textbf{First Insight (RLMC):} Hence, RLMC-CCS is equivalent to CCS if we only compute the value function corresponding to the initial starting state-action pair and ignore the other states that are visited within a path. Thus, CCS is a special case of RL Monte Carlo algorithms.

A practical challenge in applying these methods to infinite-horizon problems arises because the paths are terminated after a finite time step beyond which the simulation errors are insignificant. In the case of RLMC-CCS, for each forward simulated path, apart from the first sub-path, all the other associated sub-paths have decreasing lengths, leading to a larger bias compared to CCS when total number of simulated paths for each state-action pair is finite. In the case of finite-horizon problems, however, the RLMC-CCS algorithm will give an unbiased estimate as the paths need not be artificially terminated. 

\subsection{Temporal Difference Methods of Learning \& RLTD-CCS}
\noindent TD methods \citep{sutton1988learning} have become foundational building blocks of RL. The core idea behind TD algorithms is to update the value of the current state-action pair by simulating one or a few steps ahead, rather than waiting for the entire path to complete before making updates. Our proposed RLTD-CCS algorithm is grounded on this fundamental principle. 

To describe our approach, let us start by writing the choice-specific value function of the 1\textsuperscript{st} sub-path of the $k$\textsuperscript{th} full path in terms of the 2\textsuperscript{nd} sub-path:
\begin{equation}
    \begin{split}
    \tilde{v}^{k,1}(s_1,a_1;\hat{\theta}_u^i, \hat{\theta}_F) &= u(s_1,a_1;\hat{\theta}_u^i) + \beta\left[u(s_2,a_2;\hat{\theta}_u^i)+\gamma-\log{\left(\hat{\pi}(a_2 \vert s_2)\right)}\right]  +  \\
    &\beta^{2}\left[u(s_{3},a_{3};\hat{\theta}_u^i) + \gamma - \log{\left(\hat{\pi}(a_{3} \vert s_{3})\right)}\right] + \cdots \\
    &= u(s_1,a_1;\hat{\theta}_u^i) + \beta \left[\gamma - \log{\left(\hat{\pi}(a_2 \vert s_2)\right)}\right] + \beta \Bigg[u(s_2,a_2;\hat{\theta}_u^i) + \\
    & \beta\left[u(s_{3},a_{3};\hat{\theta}_u^i) + \gamma - \log{\left(\hat{\pi}(a_{3} \vert s_{3})\right)}\right] + \cdots \Bigg]\\
    &= u(s_1,a_1;\hat{\theta}_u^i) + \beta \left[\gamma - \log{\left(\hat{\pi}(a_2 \vert s_2)\right)}\right] + \beta \tilde{v}^{k,2}(s_2,a_2;\hat{\theta}_u^i, \hat{\theta}_F)
\end{split}
\end{equation}

The sequence of forward-simulated state-action pairs is given by $(s_1, a_1), (s_2, a_2), \ldots, (s_{T_{\text{end}}}, a_{T_{\text{end}}})$. To compute the value function for the 1\textsuperscript{st} sub-path, it is necessary to first compute the value of the 2\textsuperscript{nd} sub-path, which in turn requires accumulating the returns up to $T_{\text{end}}$. Instead of traversing till the end of the path to compute $\tilde{v}^{k,2}(s_2,a_2;\hat{\theta}_u^i, \hat{\theta}_F)$, we can replace it with the most recent value function update for the state-action pair $(s_2,a_2)$ and simplify the computation of $\tilde{v}^{k,1}(s_1,a_1;\hat{\theta}_u^i, \hat{\theta}_F)$:
\begin{equation}
    \tilde{v}^{k,1}(s_1,a_1;\hat{\theta}_u^i, \hat{\theta}_F)[\Lambda+1] = u(s_1,a_1;\hat{\theta}_u^i) + \beta \left[\gamma - \log{\left(\hat{\pi}(a_2 \vert s_2)\right)}\right] + \beta \tilde{v}(s_2,a_2;\hat{\theta}_u^i, \hat{\theta}_F)[\Lambda]
\end{equation}
In this approach, we are not explicitly accumulating the discounted rewards beyond the first simulation step but substituting it using the most recent value of $\tilde{v}(s_2,a_2;\hat{\theta}_u^i, \hat{\theta}_F)$ that was available at $[\Lambda]$ computer time. Using the general simulation step notation introduced in Equation~\ref{eq:ccs_gen_notation}, and omitting the sub-path index in the superscript for brevity, the RLTD-CCS update error (similar to as described in Equation~\ref{eq:ccs_step_wise}) can be expressed as follows:
\begin{multline}
    \Delta^{\text{\tiny{RLTD-CCS}}}_1(s,a;\hat{\theta}_u^i, \hat{\theta}_F)[\Lambda+1] = \tilde{v}^{k}(s,a;\hat{\theta}_u^i, \hat{\theta}_F)[\Lambda+1] - \tilde{v}(s,a;\hat{\theta}_u^i, \hat{\theta}_F)[\Lambda] \\
    = u(s,a;\hat{\theta}_u^i) + \beta \left[\gamma - \log{\left(\hat{\pi}(a^\prime \vert s^\prime)\right)}\right] + \beta \tilde{v}(s^\prime,a^\prime;\hat{\theta}_u^i, \hat{\theta}_F)[\Lambda] - \tilde{v}(s,a;\hat{\theta}_u^i, \hat{\theta}_F)[\Lambda]
\end{multline}
Here, the subscript “1” in $\Delta$ indicates that the update error results from a one-step forward simulation. The corresponding ``1-step'' RLTD-CCS update is then obtained by substituting this error into the learning equation presented in Equation~\ref{eq:learning_equation}:
\begin{equation}
    \tilde{v}(s,a;\hat{\theta}_u^i, \hat{\theta}_F)[\Lambda+1] = \tilde{v}(s,a;\hat{\theta}_u^i, \hat{\theta}_F)[\Lambda] + \alpha \Delta^{\text{\tiny{RLTD-CCS}}}_1(s,a;\hat{\theta}_u^i, \hat{\theta}_F)[\Lambda+1]
    \label{eq:1-step_RLTD_CCS}
\end{equation}
Essentially, we need to simulate one step forward and then use the latest entry in a choice-specific value function look-up table for the next state-action pair. Multiple updates can be done while advancing through the simulation-steps. For a simulated path, we can either update the choice-specific value functions with the most recent look-up table entries (also called \textit{on-line} learning in RL) or update the value functions in one-go after the path is terminated (also called \textit{batch} learning). The pseudocode for the on-line version of ``1-step'' RLTD-CCS is provided in the online appendix.

In Equation~\ref{eq:1-step_RLTD_CCS}, if we set the learning rate $\alpha$ to 1 and replace the choice-specific value function $\tilde{v}(s^\prime, a^\prime; \hat{\theta}_u^i, \hat{\theta}_F)$ with the ``expected'' value function, the update closely resembles the value iteration step commonly used in the inner loop of the ``polyalgorithm'' within the NFXP approach.\footnote{Although not well recognized, \cite{rust1987optimal} implemented a \textit{polyalgorithm} that initially uses value iteration to converge toward the neighborhood of the fixed point, and then transitions to ``policy iteration'' to accelerate convergence particularly when the discount factor $\beta$ is close to one \citep{rust1994structural}.} However, in the case of RLTD-CCS, we are instead using a value that has been approximated (``predicted'' in RL vocabulary) using previous forward simulations. Such approximation methods are sometimes called \textit{stochastic} Dynamic Programming in RL \citep{jaakkola1993convergence}. 

It might be useful to accumulate discounted rewards for a few steps before computing the update error. This is expected to reduce the bias, albeit at the cost of more processing time. This could be done by using the ``n-step'' RLTD-CCS algorithm. The n-step RLTD-CCS update error in terms of CCS formulation can be written as:
\begin{multline}
    \Delta_n^{\text{RLTD-CCS}}(s,a; \hat{\theta}^i_u,\hat{\theta}_F)[\Lambda+1] = u(s,a; \hat{\theta}^i_u) + \beta \left[u(s^\prime,a^\prime; \hat{\theta}^i_u) + \gamma - \log{\left(\hat{\pi}\left(a^\prime \big| s^\prime\right)\right)}\right] + \cdots \\
    \beta^{n} \left[\gamma - \log{\left(\hat{\pi}\left(a_{n+1} \big| s_{n+1}\right)\right)}\right] + \beta^n \tilde{v}(s_{n+1},a_{n+1}; \hat{\theta}^i_u,\hat{\theta}_F)[\Lambda]- \tilde{v}(s,a; \hat{\theta}^i_u,\hat{\theta}_F)[\Lambda]
\end{multline}
The n-step RLTD-CCS update rule then becomes:
\begin{equation}
    \tilde{v}(s,a;\hat{\theta}_u^i, \hat{\theta}_F)[\Lambda+1] = \tilde{v}(s,a;\hat{\theta}_u^i, \hat{\theta}_F)[\Lambda] + \alpha \Delta^{\text{\tiny{RLTD-CCS}}}_n(s,a;\hat{\theta}_u^i, \hat{\theta}_F)[\Lambda+1]
\end{equation}

\noindent \textbf{First Insight (RLTD):}  Note RLMC-CCS is a special case of n-step RLTD-CCS algorithm when $n$ is extended to the full length of the simulated path and the learning rate $\alpha$ is set according to Equation~\ref{eq:RLMC_general_form}. Hence, CCS is a special case of RLTD-CCS. 

\noindent \textbf{Second Insight (RLTD):} Similar to RLMC-CCS, in 1-step RLTD-CCS, value function updates are performed for all state-action pairs along a forward simulated path, except for the final state-action pair at the end of the path. For the n-step RLTD-CCS case, value function updates are applied to the first $T_{\text{end}} - n$ state-action pairs within the path.

\section{Monte Carlo Studies}
\noindent We assess the performance of our proposed algorithms using two Monte Carlo studies: (1) a canonical machine replacement problem (the \cite{bajari2007estimating} version of \cite{rust1987optimal}), and (2) a consumer food choice model \citep{huang2015short}.

\subsection{Machine Replacement Model} 
\noindent In each period $t=1,2,...,T\leq \infty$, a decision-making superintendent observes a discrete state $s_t \in S = \left\lbrace1, 2, ..., N \right\rbrace$ of a machine. She also observes certain information about the machine (e.g., component failure, technician's report, batch, etc.) denoted by $\epsilon_{t}$ that is unobserved to the researcher. After observing $\left\lbrace s_{t}, \epsilon_{t} \right\rbrace$, she decides whether to replace the machine ($a_t=1$) or not ($a_t=0$). The superintendent has the following beliefs about the evolution of the machine's state conditional on the action taken: 
\begin{equation}
    s_{t^\prime} = \begin{cases}
    \min\left\lbrace s_t + 1, N \right\rbrace , & \text{if $a_t=0$} \\
    1 , & \text{if $a_t=1$}
    \end{cases}.
    \label{eq:mrm_state_evolution}
\end{equation}
Thus, when a machine is replaced its state is reset to 1 and the machine starts anew. This is referred to as the \textit{regenerative optimal stopping property} of the stochastic process by \cite{rust1987optimal}. Also, $N$ is the absorbing state in this specification, that is, once this state is reached it doesn't change until an action to replacement is taken.

Each action  has an associated cost with the cost of replacement ($a_t=1$) assumed to be larger than that of maintenance ($a_t=0$), with the probability of failure increasing with usage. The current utility due to each action is given by:
\begin{equation}
	U(s_t, \epsilon_t(a), a; \theta_u) = \begin{cases}
	-c(s_t, \theta_{\text{MC}})+\epsilon_t(0), & a_t = 0 \\
	-\theta_{\text{RC}}+\epsilon_t(1), & a_t = 1
	\end{cases}, 
\end{equation}
where, $\theta_u = (\theta_{\text{MC}}, \theta_{\text{RC}})$, $c(\cdot)$ is some cost function for the observed state $s_t$ and maintenance cost parameter $\theta_{\text{MC}}$, and $\theta_{\text{RC}}$ is the replacement cost parameter.

The superintendent's objective is to minimize the present discounted value (PDV) of the stream of costs. The value function can be written as:
\begin{equation}
	V(s_t, \epsilon_t; \theta_u) = \max_{a_t\in\{0,1\}} \left[ V(s_t, \epsilon_t(a), a_t; \theta_u) \right],
\end{equation}
Writing in terms of the choice-specific value function, we get:
\begin{equation}
	V(s_t, \epsilon_t(a), a_t; \theta_u) = \begin{cases}
	-c(s_t, \theta_\text{MC})+\epsilon_t(0)+\beta \mathbb{E} \left[V(s_{t^\prime}, \epsilon_{t^\prime};\theta_u) \big| s_t, \epsilon_t(0), a_t=0\right] , & a_t=0 \\
	-\theta_\text{RC}+\epsilon_t(1) + \beta\mathbb{E} \left[V(1, \epsilon_{t^\prime};\theta_u) \big| s_t, \epsilon_t(1), a_t=1\right], & a_t=1
	\end{cases} .
\end{equation}
The deterministic component of the choice-specific function can then be written as:
\begin{equation}
	v(s_t, a_t; \theta_u) = \begin{cases}
	-c(s_t, \theta_\text{MC})+\beta \mathbb{E} \left[V(s_{t^\prime}, \epsilon_{t^\prime};\theta_u) \big| s_t, \epsilon_t(0), a_t=0\right] , & a_t=0\\
	-\theta_\text{RC}+\epsilon_t(1) + \beta\mathbb{E} \left[V(1, \epsilon_{t^\prime};\theta_u) \big| s_t, \epsilon_t(1), a_t=1\right], & a_t=1
	\end{cases} .
\end{equation}
Assuming Type-1 MEV distribution for the unobserved error terms, we can express the CCPs for each state-action pair in terms of choice-specific value functions as :
\begin{equation}
	\pi(a_t \vert s_t; \theta_u) = \begin{cases}
	\dfrac{e^{v(s_t,0; \theta_u)}}{e^{v(s_t,0; \theta_u)} + e^{v(s_t,1; \theta_u)}}, & a_t=0 \\
 \\
	\dfrac{e^{v(s_t,1; \theta_u)}}{e^{v(s_t,0; \theta_u)} + e^{v(s_t,1; \theta_u)}}, & a_t=1
	\end{cases}.
    \label{eq:mrmr_ccp}
\end{equation}

The estimation using the CCS and RL algorithms is done in two steps: (1) directly estimating the transition and choice probabilities from the (synthetic) data (which is the same for all algorithms), followed by (2) a value function computation (forward simulation) step which is embedded within an outer optimization loop to estimate the structural parameters $\theta_u$ using the MDE procedure.

In this Monte Carlo study, we used a small state-space machine replacement model with the maximum number of states, $N$, equal to 5 and two actions at the superintendent's disposal in every period, either maintenance or replacement of the machine. The size of the state-action space was $N \times |\mathcal{A}| = 10$.  The state transitions followed the rule specified in Equation~\ref{eq:mrm_state_evolution}. The costs associated with the actions were fixed to $\Bar{\theta}_u = (\Bar{\theta}_\text{MC}, \Bar{\theta}_{RC}) = (1, 4)$. The discount factor $\beta$ was assumed to be $.90$. The choice-specific value-functions, $v(s_t, a_t; \Bar{\theta}_{u})$, were computed using the fixed-point algorithm (the inner loop of NFXP \citep{rust1987optimal}). The choice-specific value functions were then input into Equation~\ref{eq:mrmr_ccp} to obtain the corresponding CCPs. Then using the transition probabilities and the computed CCPs, a synthetic dataset for 10,000 machines over a period of 100 weeks was generated through the Monte Carlo procedure.

Given the objective of accelerating value function computation and brevity of space, our discussion below is focused on the second step as that is the main difference between the CCS and RL-based estimation procedures. The outer parameter search loop was kept common across all the algorithms. For computing the choice-specific value functions, we pre-simulated a set of paths using the already estimated transition and choice probabilities.  Each path was terminated after a period $T_{\text{end}}$. Since our RL-based algorithms can update the value functions intermittently, often every time a state-action pair is visited in a path, we wanted to test how the choice of $T_{\text{end}}$ affects the accuracy and speed of estimation.
We varied $T_{\text{end}}$ from 4 to 200, and for each termination length, we pre-simulated 50 sets of 500 paths ($N_{\text{path}}$). All computations were performed on a university's High Performance Computing (HPC) cluster. For every individual estimation process,  we allocated a single CPU core. We avoided parallelization to compare the default computation speeds. For each unique $T_{\text{end}}$ case, we estimated the model for all the 50 sets of paths separately. From these runs, we computed the mean, standard deviation, and root mean square error (RMSE) of the parameter estimates. The RMSE was calculated relative to the modeled structural utility, \( \bar{\theta}_u \). Additionally, we tracked the number of times the outer optimizer evaluated the distance (measured using the Euclidean norm or \( \ell^2 \)-norm) between the estimated and predicted CCPs (denoted as ``\#fevals'') and recorded the final minimized distance at convergence.

As expected, the number of value functions updates during a single inner loop execution was much higher for the RL algorithms. For example, when $T_{\text{end}}$ was 50, for each guess $\hat{\theta}_u$, the value functions corresponding to each state-action pair were updated roughly 50 times in CCS. In 1-step RLTD-CCS, the choice-specific value function for the state-action pair (1,0) was updated more than 8000 times, for (1,1) more than 1200 times, and so on for each optimization loop (the online appendix provides more details).

\begin{figure}[htp]
    \centering
    \includegraphics[width=1\linewidth]{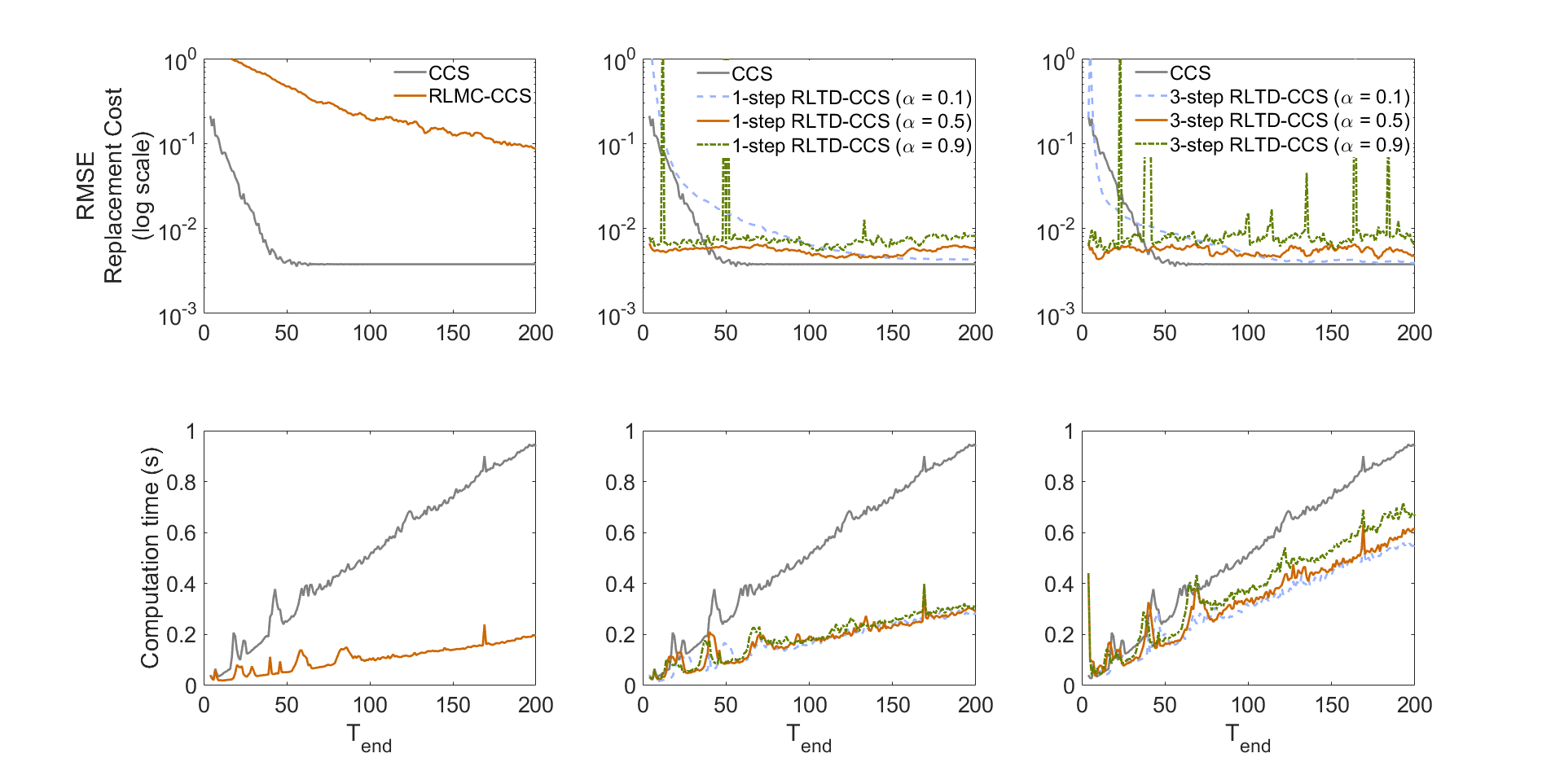}
    \caption{\centering Plots showing (\textit{top row}) the estimation accuracy (RMSEs for the estimated replacement cost) and (\textit{bottom row}) the mean computation times when forward simulated paths of different lengths ($T_{\text{end}}$) are used in different algorithms.}
    \label{fig:plot_est_psa_data}
\end{figure}

The accuracy of the estimates and computation times are illustrated in Figure \ref{fig:plot_est_psa_data}. As is evident from the top row of the figure, for CCS, the RMSEs of the replacement cost parameter ($\hat{\theta}_\text{RC}$) improved if the simulated paths were longer, reaching the best possible values around a $T_{\text{end}}$ equal to 50. Simulating and using longer paths did not improve the accuracy of the estimates. Interestingly, the literature (e.g. \cite{hotz1994simulation}), uses simulated path lengths with a $T_{\text{end}}$ equal to 50 which seems appropriate based on our findings.  

In the case of RLMC-CCS, the estimation accuracy was limited. This is to be expected, as in the infinite-horizon case, sub-paths with decreasing termination lengths introduce larger errors, leading to less accurate estimates.

We provide results for 1-step and 3-step RLTD-CCS algorithms for three different values of step-size parameters ($\alpha$ equal to .1, .5, and .9). When $\alpha$ was equal to .5 and .9, the RMSE of the replacement cost parameter reached its best value even for much shorter paths ($T_{\text{end}}$ less than or equal to 10). This was a significant improvement over CCS. For the smaller step-size parameter case ($\alpha$ equal to .1), the RMSE improvement with the increasing length of the simulated paths was gradual. This is as expected as a smaller $\alpha$ meant a smaller weight was given to every update. However, this resulted in smoother convergence. On the other hand, when $\alpha$ was equal to .9, the RMSE increased for several path lengths, indicating that the estimation routine was not able to converge for these lengths. Using even larger step-size parameter values resulted in nosier and poorer convergence. The mid-range $\alpha$ of .5 performed the best so we make it the default for the rest of the analysis.

In the bottom row of the figure, we've plotted the total mean computation times. RLMC-CCS was the fastest for most $T_\text{end}$ cases. This was because the optimizer was not able to minimize the cost function beyond a few iterations (due to poor estimates) and exited faster than the other algorithms. For the smaller $T_{\text{end}}$ cases, the computation times for CCS, 1-step RLTD-CCS, and 3-step RLTD-CCS were similar. However, for the cases when the simulated paths were longer, CCS took longer time. Between 1-step and 3-step RLTD-CCS, 1-step RLTD-CCS was faster. This is as expected, as more processing steps are involved while computing the value functions in the inner loop of the 3-step version of RLTD-CCS.
\begin{table}[htp]
\captionsetup{skip=2pt}
\caption{Estimates for the Machine Replacement Model}
\label{tab:mrm_est_results}
    \scriptsize
    \centering
    \begin{tabular}{l l | c c c c}
        \multicolumn{6}{c}{\scriptsize{$(\Bar{\theta}_\text{MC},\Bar{\theta}_\text{RC}) = (1,4)$, $\beta = .90$}, \scriptsize{state-action space size = 10, $N_{\text{path}} = 500$}} \\
        \toprule
        \multicolumn{2}{l|}{Parameter} & CCS & RLMC-CCS & 1-step RLTD-CCS & 3-step RLTD-CCS \\
        {} & {} & {} & {} & {$\alpha = .5$} & {$\alpha=.5$} \\
        \midrule
        \vspace{0.1cm}
        {$T_{\text{end}}$} & {} & {10} & {10} & {10} & {10} \\
        {$\theta_\text{MC}$} & {Mean (Std.)} & {1.0011 (2.04E-02)} & {.7341 (1.89E-02)} & {.9999 (6.38E-04)} & {.9995 (8.03E-04)} \\
        \vspace{0.1cm}
        {} & {RMSE} & {2.03E-02} & {.2665} & {6.39E-04} & {9.45E-04} \\
        {$\theta_\text{RC}$} & {Mean (Std.)} & {4.0084 (9.16E-02)} & {2.9026 (8.07E-02)} & {4.0046 (2.96E-03)} & {4.0017 (4.05E-03)} \\
        \vspace{0.1cm}
        {} & {RMSE} & {9.11E-02} & {1.1002} & {5.43E-03} & {4.36E-03} \\
        \vspace{0.1cm}
        {$\ell^2$-norm} & {Mean (Std.)} & {1.24E-02 (5.02E-03)} & {4.65E-02 (1.21E-02)} & {1.16E-03 (1.45E-04)} & {1.17E-03 (2.05E-04)} \\
        \vspace{0.1cm}
        {\#fevals} & {Mean (Std.)} & {66.6 (7.52)} & {60.6 (6.06)} & {82.74 (10.60)} & {84.48 (12.93)} \\
        \vspace{0.1cm}
        {Time (s)} & {Mean (Std.)} & {3.92E-02 (4.83E-03)} & {2.13E-02 (7.07E-03)} & {2.57E-02 (3.86E-03)} & {7.06E-02 (1.63E-02)} \\
        \midrule
        \vspace{0.2cm}
        {$T_{\text{end}}$} & {} & {50} & {50} & {50} & {50} \\
        {$\theta_\text{MC}$} & {Mean (Std.)} & {.9998 (4.57E-04)} & {.8798 (3.49E-02)} & {.9999 (7.81E-04)} & {.9996 (1.06E-03)} \\
        \vspace{0.1cm}
        {} & {RMSE} & {5.01E-04} & {.1251} & {7.81E-04} & {1.13E-03} \\
        {$\theta_\text{RC}$} & {Mean (Std.)} & {4.0035 (2.05E-03)} & {3.5549 (.1516)} & {4.0046 (3.86E-03)} & {4.0027 (5.28E-03)} \\
        \vspace{0.1cm}
        {} & {RMSE} & {4.09E-03} & {.4697} & {6E-03} & {5.91E-03} \\
        \vspace{0.1cm}
        {$\ell^2$-norm} & {Mean (Std.)} & {1.14E-03 (1.4E-04)} & {3.60E-02 (1.02E-02)} & {1.16E-03 (1.35E-04)} & {1.27E-03 (1.93E-04)} \\
        \vspace{0.1cm}
        {\#fevals} & {Mean (Std.)} & {85.56 (8.20)} & {52.02 (4)} & {79.68 (10.45)} & {88.58 (11.85)} \\
        \vspace{0.1cm}
        {Time (s)} & {Mean (Std.)} & {.2492 (2.94E-02)} & {5.23E-02 (8.76E-03)} & {8.65E-02 (1.77E-02)} & {.1655 (2.63E-02)} \\
        \bottomrule
    \end{tabular}
\end{table}

Table~\ref{tab:mrm_est_results} includes the results for estimation runs when $T_{\text{end}}$ was equal to 10 and 50. For $T_{\text{end}}$ equal to 10, the RMSEs using CCS were one to two orders of magnitude worse compared to RLTD-CCS. Additionally, the minimized $\ell^2$-norm values indicate that the predicted CCPs using RLTD-CCS methods better matched the estimated CCPs with values roughly an order of magnitude smaller than those from CCS. This suggests a closer fit to the data.

This improvement is also reflected in the number of function evaluations required before the optimizer exits. On average, RLMC-CCS exits after approximately 60 iterations, CCS after 66, while RLTD-CCS variants typically run over 80 iterations, indicating a better convergence behavior.

For longer paths ($T_{\text{end}}$ equal to 50), the differences narrow significantly. CCS and RLTD-CCS algorithms achieve similar RMSEs, $\ell^2$-norms, and optimizer iteration counts. These results suggest that RLTD-CCS can achieve high estimation accuracy even with shorter simulation paths without sacrificing performance.

Comparing the computation times for CCS ($T_{\text{end}}$ equal to 50) and 1-step RLTD-CCS  ($T_{\text{end}}$ less than or equal to 10), the 1-step RLTD-CCS reached the same accuracy compared to CCS but in 1/10\textsuperscript{th} time. 

In sum, the RL based methods are as accurate as CCS but require shorter sample paths and less memory for the canonical machine replacement model. The next section investigates these properties for a large state-action space model.

\subsection{Food Choice Model}
\noindent To evaluate the performance of our algorithms in large state-action space problems, we introduce an infinite-horizon consumer food choice model that builds on \cite{huang2015short}.\footnote{This problem is inspired by a separate project with an online recipe box retailer. For alternative models of demand for food products see e.g., \cite{dube2004multiple}, \cite{allcott2019food}..} A consumer, in every period $t = 1,2,...,T \leq \infty$, chooses between $M$ different mutually exclusive recipes from an online platform for home cooking. The action $a_{t} = m$ if the recipe $m$ is chosen from the online menu and zero if the individual foregoes ordering ($a_t = 0$).

Each recipe varies in the content of three attributes, salt ($SLT_m$), sugar ($SUG_m$), and saturated fat ($SAT_m$). For the outside option ($a_t=0$), these attributed are assumed zero. Based on previous consumption decisions the consumer accumulates stocks of these attributes given by ${SLT}_t^{\text{stock}}$, ${SUG}_t^{\text{stock}}$ and ${SAT}_t^{\text{stock}}$ respectively. We impose a maximum limit to the stock values ($STOCK_\text{max}$) of these attributes. The dynamics of the stock variables are given by:
\begin{equation}
    \begin{split}
        {SLT}_{t^\prime}^{\text{stock}} &= \min\left\lbrace \left((1-\delta_0)*\mathds{1}\left\lbrace a_t=0\right\rbrace +(1-\delta_1)*\mathds{1} \left\lbrace a_t = m \right\rbrace \right)SLT^{\text{stock}}_t+SLT_m, STOCK_\text{max}\right\rbrace,\\
        {SUG}_{t^\prime}^{\text{stock}} &= \min\left\lbrace \left((1-\delta_0)*\mathds{1}\left\lbrace a_t=0\right\rbrace +(1-\delta_1)*\mathds{1} \left\lbrace a_t = m \right\rbrace \right)SUG^{\text{stock}}_t+SUG_m, STOCK_\text{max}\right\rbrace,\\
        {SAT}_{t^\prime}^{\text{stock}} &= \min\left\lbrace \left((1-\delta_0)*\mathds{1}\left\lbrace a_t=0\right\rbrace +(1-\delta_1)*\mathds{1} \left\lbrace a_t = m \right\rbrace \right)SAT^{\text{stock}}_t+SAT_m, STOCK_\text{max}\right\rbrace,\\
    \end{split}
\end{equation}
where $\delta_0$ and $\delta_1$ denote the stock depreciation factors when the consumer forgoes ordering and when they order a recipe from the menu, respectively \citep{grossman1972health,benkard2000learning}. In order to allow for a regenerative property to make it comparable to the machine replacement model as in 
\cite{rust1987optimal}, we set $\delta_0 = 1$ (complete depreciation). This implies that that the stock levels reset to zero if the consumer skips ordering. We set $\delta_1 = 0$ (no depreciation), meaning that consumers fully carry over their previous stock levels when they place an order. In principle, as in \cite{grossman1972health}, $\delta_{0}$ and $\delta_{1}$ can be equal taking a value between 0 and 1.

Given parameters $\theta_{SLT}, \theta_{SUG}, \theta_{SAT}$, the disutility, $r_{t}^{\text{stock}}$, from the accumulated stocks of these attributes is expressed as:
\begin{equation}
    r_{t}^{\text{stock}} = -\theta_{SLT} {SLT}_t^{\text{stock}} - \theta_{SUG} {SUG}_t^{\text{stock}} - \theta_{SAT} {SAT}_t^{\text{stock}}
\end{equation}

Each recipe is assigned a fixed parameter, $r_m^{\text{fixed}}$, capturing intrinsic recipe-specific preferences such as cuisine type, dish category, and other non-nutritional characteristics. We assume these parameters are known to the researcher. Here we focus solely on the nutritional attributes of recipes.

To model variety-seeking behavior, where consumers derive disutility from repeated selections of the same recipe across consecutive periods, we introduce the state variable $h_t^{\text{variety}}$. Its evolution is defined by:
\begin{equation}
    h_{t+1}^{\text{variety}} = \begin{cases}
        \min\left\lbrace h_t^{\text{variety}} + 1, H_{\text{max}} \right\rbrace & \text{ if } a_{t} = a_{t-1}, \\
        0 & \text{ otherwise},
    \end{cases}
\end{equation}
where, $H_{\text{max}}$ is the maximum level of accumulated disutility from repeated choices. Given a parameter $\theta_{\text{variety}}$, the disutility from lack of variety at time $t$ is specified as:
\begin{equation}
    r_t^{\text{variety}} = -\theta_{\text{variety}} h_t 
\end{equation}

Foregoing ordering leads to a disutility which is assumed to be fixed at $-\theta_{\text{skip}}$. This could be imagined as time spent in arranging ingredients and limited preference for self-assembled and cooked food, etc. For simplicity, skipping consumption resets the stock variables.

The single-period utility received in period $t$ is given by:
\begin{equation}
    U(s_t,\epsilon_{t}(a_t),a_t; \theta_u) = 
    \begin{cases}
        r_m^{\text{fixed}} + r_{t}^{\text{stock}} +  r_t^{\text{variety}} + \epsilon_{t}(m), & \text{if } \; a_t=m, \\
        -\theta_{\text{skip}} + \epsilon_{t}(0)& \text{if } \; a_t=0,
    \end{cases}
\end{equation}
where, $\epsilon_{t}(a_t)$ is the random utility component (unoberved by the researcher) associated with the action $a_t$, and $\theta_u = (\theta_{SLT}, \theta_{SUG}, \theta_{SAT}, \theta_{\text{variety}}, \theta_{\text{skip}})$ are the structural utility parameters of interest. The consumer's objective is to maximize the PDV of the stream of utility. 

The observable state variables for the consumer's problem are:
\begin{equation}
    s_t = ({SLT}_t^{\text{stock}}, \; {SUG}_t^{\text{stock}},  \; {SAT}_t^{\text{stock}}, \; h_t^{\text{variety}})
\end{equation}

The deterministic component of the choice-specific function can then be written as:
\begin{equation}
    v(s_t,a_t;\theta_u) =  
    \begin{cases}
        r_m^{\text{fixed}} + r_{t}^{\text{stock}} +  r_t^{\text{variety}} + \beta \mathbb{E} \left[V(s_{t^\prime}, \epsilon_{t^\prime}; \theta_u) \big| s_{t}, \epsilon(a_t), a_{t}=m\right] \\
        -\theta_{\text{skip}} + \beta \mathbb{E} \left[V(s_{t^\prime}, \epsilon_{t^\prime}; \theta_u) \big| s_{t}, \epsilon(a_t), a_{t}=0\right]
    \end{cases}
\end{equation}
Hence, the choice-probability for the selecting the $m$\textsuperscript{th} recipe in period $t$ is:
\begin{equation}
    \pi(a_{mt} \big| s_t; \theta_u) = \dfrac{e^{v(s_t,a_t=m; \theta_u)}}{\sum_{a_t=0}^{a_t=M} e^{v_j (s_t, a_t;\theta_u)} }
\end{equation}

The synthetic data generation procedure for the food choice model is provided in the online appendix. We estimated eight versions of the food choice model, with state-action space ranging from 3,891 to 5.77 million (see Table~\ref{tab:fcm_model_spec} for details on each version). For each version of the food choice model, we first generated 50 sets of forward simulation paths, with each set containing paths of similar lengths ($T_{\text{end}}$ equal to 5, 10 and 50). For each version of the model, the number of forward simulated paths was roughly kept 20 to 25 times the size of the state-action space. We randomized the starting states of each path by drawing with replacement from the observed states seen in the synthetic dataset. In the second stage, we estimated the structural parameters using CCS and 1-step RLTD-CCS algorithm. We also estimated a few smaller state-action space food choice models using 3-step RLTD-CCS. However, as estimating these took longer than the 1-step RLTD-CCS without substantial improvement in accuracy, we have not included these results here. For RLTD-CCS, $\alpha$ was fixed to .5.
\begin{table}[!ht]
\captionsetup{skip=2pt}
\caption{Food Choice Model Specifications}
\label{tab:fcm_model_spec}
\tiny
   \centering
    \begin{tabular}{c | c | c |c c c c c | c | c | c}
        \toprule
        Model & $STOCK_{\text{max}}$ & $H_{\text{max}}$ & \multicolumn{5}{c|}{Modeled Structural Parameter Values} & $\beta$ & M & State-action \\
         & & & $\Bar{\theta}_{SLT}$ & $\Bar{\theta}_{SUG}$ & $\Bar{\theta}_{SAT}$ & $\Bar{\theta}_{\text{variety}}$ & $\Bar{\theta}_{\text{skip}}$ & & & space size \\
         \midrule
         Case 1a & 3 & 3 &  .5 &  .5 &  .75 &  .1 &  5 & .90 & 2 & 3,891 \\
         Case 1b & 3 & 3 &  .5 &  .5 &  .75 &  .1 &  5 & .90 & 10 & 71,291 \\
         Case 2a & 5 & 5 &  .5 &  .5 &  .75 &  .1 &  9 & .90 & 2 & 30,001 \\
         Case 2b & 5 & 5 &  .5 &  .5 &  .75 &  .1 &  9 & .90 & 10 & 550,011 \\
         Case 3a & 6 & 6 &  .5 &  .5 &  .75 &  .1 &  12 & .90 & 2 & 62,211 \\
         Case 3b & 6 & 6 &  .5 &  .5 &  .75 &  .1 &  12 & .90 & 10 & 1.14 mil. \\
         Case 4a & 9 & 9 &  .5 &  .5 &  .75 &  .1 &  15 & .90 & 2 & 314,931 \\
         Case 4b & 9 & 9 &  .5 &  .5 &  .75 &  .1 &  15 & .90 & 10 & 5.77 mil. \\
        \bottomrule
    \end{tabular}
\end{table}

The estimation results are provided in Table~\ref{tab:fcm_1a}--\ref{tab:fcm_4b}. When very short paths were used ($T_{\text{end}}$ equal to 5), 1-step RLTD-CCS achieved substantially better performance than CCS across most models. Specifically, RLTD-CCS estimated the utility parameter $\theta_{u}$ with RMSE values that were three to four orders of magnitude smaller than those obtained using CCS. The $\ell^2$-norm values exhibit a similar pattern, where the minimized values using CCS were approximately three orders of magnitude larger than those from RLTD-CCS. This indicates a significantly poorer fit to the observed conditional choice probabilities. Additionally, the number of function evaluations required for convergence suggests that the optimizer typically performs about two and a half times more iterations under RLTD-CCS than under CCS, reflecting the optimizer's ability to more effectively minimize the objective.

When long paths were used ($T_{\text{end}}$ equal to 50), CCS achieved estimation accuracy for the utility parameters that was roughly comparable to RLTD-CCS (which remained about the same for all the three path lengths). However, the corresponding $\ell^2$-norm values remained about an order of magnitude larger, which suggests that CCS may require even longer forward-simulated paths to reduce estimation bias. This interpretation is further supported by the slightly lower number of function evaluations observed for CCS, which indicates that the optimization routine often terminates earlier before reaching a better minimum.

These results are consistent with observations from the machine replacement model, where achieving comparable RMSE performance with CCS required forward-simulated paths that were approximately ten times longer than those used by RLTD-CCS.

Generating and storing longer paths is both time consuming and memory intensive. For example, in the case of the smallest state-space model (Model: `Case 1a' with a state-action space size of 3,891), generating a set of $25 \times 3891$ paths with a $T_{\text{end}}$ equal to 5 took around 2 seconds and a memory space of roughly 385 kilobytes was required to store the paths. On the other hand, generating paths with a $T_{\text{end}}$ equal to 50 roughly took 22 seconds and the paths occupied about 4.2 megabytes of memory space. While these numbers might look trivial in the case of the small state-space model, when the state-space increased the speed and memory problem became exponentially challenging. For example, in the case of the largest state-space models, the path generation process took several hours and had to be done on multiple computing cores to avoid runtime memory issues. Considering memory requirements, for the model with the state-space size of 1.14 million, paths with a $T_{\text{end}}$ equal to 5 required almost 310 megabytes of memory, while paths with a $T_{\text{end}}$ equal to 50 required almost 3.1 gigabytes of memory. The memory figures quoted here are when the path data was stored in the most compressed, MATLAB compatible `.mat' format. Storing data as in a more commonly used `comma-separated value' or `.csv' format takes about 3.5 to 4 times more memory.

CCS was generally faster than the 1-step RLTD-CCS for the shorter path lengths. However, this was at the expense of a poorer estimation accuracy. In a few cases, CCS even estimated the wrong sign for the $\theta_{\text{variety}}$ parameter. The mean of the estimated $\theta_{\text{variety}}$ parameter was -.0002 and -.0455 for models with state-action space size of 30,001 (Table~\ref{tab:fcm_2a}) and 62,211 (Table~\ref{tab:fcm_3a}) respectively. On the other hand, using 1-step RLTD-CCS, the mean of the estimates were +.1000 and +.0999, which were extremely close to the modeled value ($\bar{\theta}_{\text{variety}}$) of +.1. 

In the case of the largest state-action space size model (5.77 million states, Table~\ref{tab:fcm_4b}), the estimation routines with longer paths ($T_{\text{end}}$ equal to 50) were terminated by the HPC scheduler as these overran the allowed 24 hours limit. While these could be estimated in principle on a different cluster/PC where the time limits are more relaxed, it provided a good indication of the practical limits. As we were able to estimate the same model comfortably within the time limit using 1-step RLTD-CCS on shorter paths, the RL algorithm provide a higher ceiling on the state-action space size of the models that could be easily estimated. 

\begin{figure}[h]
    \centering
    \includegraphics[width=0.4\linewidth]{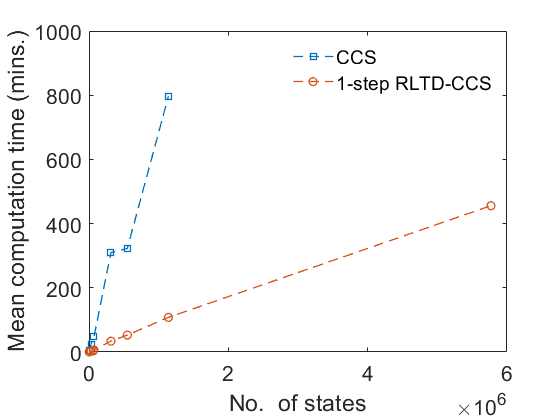}
    \caption{\centering Mean computation times vs state-action space size for CCS and 1-step RLTD-CCS algorithms.}
    \label{fig:comp_time_vs_state_space}
\end{figure}

Comparing the mean estimation computation times for CCS (when $T_{\text{end}}$ was equal to 50) with 1-step RLTD-CCS (when $T_{\text{end}}$ was equal to 5), RLTD-CCS was found to be 6 to 14 times faster than the CCS algorithm across the eight models. The computation times for both the algorithms are graphically illustrated in Figure \ref{fig:comp_time_vs_state_space}. The much steeper slope of CCS compared to RLTD-CCS underscores the latter's superior computational performance in estimating larger state-action space models in a fraction of the time without using large amounts of memory.

\begin{table}[!ht]
\captionsetup{skip=2pt}
\caption{Estimates for the Food Choice Model (Case 1a)}
\label{tab:fcm_1a}
\tiny
\setlength{\tabcolsep}{1.8pt}
   \centering
    \begin{tabular}{l l | c c | c c | c c}
        \multicolumn{8}{c}{\scriptsize{$(\Bar{\theta}_{SLT},\Bar{\theta}_{SUG},\Bar{\theta}_{SAT},\Bar{\theta}_{\text{variety}},\Bar{\theta}_{\text{skip}}) = ( .5, .5, .75, .1, 5)$, $\beta = .90$}}\\
        \multicolumn{8}{c}{\scriptsize{no. of recipes ($M$) = 2, state-action space size = 3,891}}\\
        \toprule
        \multicolumn{2}{l|}{Parameter} & CCS & 1-step RLTD-CCS & CCS & 1-step RLTD-CCS & CCS & 1-step RLTD-CCS \\
        \midrule
        \vspace{0.1cm}
        $T_{\text{end}}$ & & 5 & 5 & 10 & 10 & 50 & 50\\
        $\theta_{SLT}$ & Mean (Std.) &  .4293 (7.43E-03) &  .5000 (6.23E-05) &  .4711 (4.91E-03) &  .5000 (5.55E-05) &  .4999 (9.29E-05) &  .5000 (5.82E-05)\\
        \vspace{0.1cm}
         & RMSE & 7.10E-02 & 8.11E-05 & 2.92E-02 & 6.12E-05 & 9.44E-05 & 6.39E-05\\
        $\theta_{SUG}$ & Mean (Std.) &  .4379 (6.71E-03) &  .5000 (4.58E-05) &  .4709 (5.18E-03) &  .5000 (4.88E-05) &  .4999 (9.92E-05) &  .5000 (4.18E-05)\\
        \vspace{0.1cm}
        & RMSE & 6.23E-02 & 4.55E-05 & 2.94E-02 & 5.11E-05 & 9.85E-05 & 4.14E-05\\
        $\theta_{SAT}$ & Mean (Std.) &  .6610 (8.51E-03) & .7500 (4.71E-05) & .7080 (5.83E-03) & .7500 (4.87E-05) & .7499 (1.13E-04) & .7500 (6.41E-05) \\
        \vspace{0.1cm}
        & RMSE & 8.93E-02 & 4.68E-05 & 4.23E-02 & 5.21E-05 & 1.12E-04 & 6.49E-05\\
        $\theta_{\text{variety}}$ & Mean (Std.) & .0464 (6.43E-03) & .0999 (3.10E-05) & .075 (5.92E-03) & .1000 (4.42E-05) & .1000 (7.69E-05) & .1000 (3.75E-05)\\
        \vspace{0.1cm}
        & RMSE & 5.39E-02 & 3.21E-05 & 2.55E-02 & 4.79E-05 & 7.62E-05 & 3.75E-05\\
        $\theta_{\text{skip}}$ & Mean (Std.) & 3.4280 (5.46E-02) & 4.9999 (5.18E-04) & 4.5060 (4.88E-02) & 5.0003 (5.08E-04) & 4.9998 (8E-04) & 5.0001 (5.07E-04)\\
        \vspace{0.1cm}
        & RMSE & 1.5729 & 5.14E-04 & .4963 & 5.84E-04 & 8.09E-04 & 5.31E-04\\
        \vspace{0.1cm}
        $\ell^2$-norm & Mean (Std.) & 1.03889 (2.30E-02) & 2.21E-03 (1.46E-03) & .7036 (1.70E-02) & 1.36E-03 (1.40E-03) & 1.10E-02 (3.33E-04) & 1.33E-03 (1.34E-03)\\
        \vspace{0.1cm}
        \#fevals & Mean (Std.) & 136.56 (3.10) & 307.44 (14.82) & 143.76 (4.68) & 313.44 (11.66) & 260.76 (8.84) & 313.44 (12.21)\\
        \vspace{0.1cm}
        Time (m) & Mean (Std.) & .1531 (1.7E-02) & .2785 (3.03E-02) & .3081 (2.91E-02) & .3771 (3.38E-02) & 3.75 (.2323) & 1.2715 (.1276)\\
        \bottomrule
    \end{tabular}
\end{table}
\begin{table}[!ht]
\captionsetup{skip=2pt}
\caption{Estimates for the Food Choice Model (Case 1b)}
\label{tab:fcm_1b}
\tiny
\setlength{\tabcolsep}{1.8pt}
   \centering
    \begin{tabular}{l l | c c | c c | c c}
        \multicolumn{8}{c}{\scriptsize{$(\Bar{\theta}_{SLT},\Bar{\theta}_{SUG},\Bar{\theta}_{SAT},\Bar{\theta}_{\text{variety}},\Bar{\theta}_{\text{skip}}) = (.5,.5,.75,.1,5)$, $\beta = .90$}}\\
        \multicolumn{8}{c}{\scriptsize{no. of recipes ($M$) = 10, state-action space size = 71,291}}\\
        \toprule
        \multicolumn{2}{l|}{Parameter} & CCS & 1-step RLTD-CCS & CCS & 1-step RLTD-CCS & CCS & 1-step RLTD-CCS \\
        \midrule
        \vspace{0.1cm}
        $T_{\text{end}}$ & & 5 & 5 & 10 & 10 & 50 & 50\\
        $\theta_{SLT}$ & Mean (Std.) & .3530 (4.70E-03) & .5000 (6.46E-05) & .4183 (3.33E-03) & .5000 (5.74E-05) & .4999 (9.15E-05) & .5000 (6.49E-05)\\
        \vspace{0.1cm}
         & RMSE & .1470 & 6.91E-05 & 8.16E-02 & 7.09E-05 & 9.54E-05 & 7.49E-05\\
        $\theta_{SUG}$ & Mean (Std.) & .2984 (4.14E-03) & .5000 (6.49E-05) & .4008 (3.47E-03) & .5000 (6.04E-05) & .4999 (7.55E-05) & .5000 (7.19E-05)\\
        \vspace{0.1cm}
        & RMSE & .2015 & 7.07E-05 & 9.92E-02 & 6.29E-05 & 8.47E-05 & 7.25E-05\\
        $\theta_{SAT}$ & Mean (Std.) &  .5932 (3.62E-03) & .7500 (5.95E-05) & .6683 (3.90E-03) & .7500 (6.04E-05) & .7499 (6.93E-04) & .7500 (5.33E-05) \\
        \vspace{0.1cm}
        & RMSE & .1567 & 5.89E-05 & 8.17E-02 & 6.26E-05 & 7.53E-05 & 5.29E-05\\
        $\theta_{\text{variety}}$ & Mean (Std.) & .0303 (3.49E-03) & .1000 (5.68E-05) & .065 (2.96E-03) & .1000 (7.23E-05) & .1000 (6.99E-05) & .1000 (6.87E-05)\\
        \vspace{0.1cm}
        & RMSE & 6.97E-02 & 5.63E-05 & 3.49E-02 & 7.27E-05 & 6.93E-05 & 6.84E-05\\
        $\theta_{\text{skip}}$ & Mean (Std.) & 1.7972 (3.32E-02) & 5.0002 (7.31E-04) & 3.5601 (3.86E-02) & 5.0004 (6.30E-04) & 4.9994 (9.65E-04) & 5.0002 (7.35E-04)\\
        \vspace{0.1cm}
        & RMSE & 3.2029 & 7.61E-04 & 1.440 & 7.45E-04 & 1.1E-03 & 7.64E-04\\
        \vspace{0.1cm}
        $\ell^2$-norm & Mean (Std.) & .7255 (6.07E-02) & 3.98E-04 (2.88E-04) & .5764 (5.36E-03) & 3.95E-04 (2.95E-04) & 9.65E-03 (1.15E-04) & 3.93E-04 (2.93E-04)\\
        \vspace{0.1cm}
        \#fevals & Mean (Std.) & 138 (4.19) & 328.8 (14.99) & 141.48 (5.15) & 329.64 (17.62) & 231.84 (5.65) & 331 (17.48)\\
        \vspace{0.1cm}
        Time (m) & Mean (Std.) & 3.44 (.21) & 6.25 (.66) & 6.55 (.38) & 8.60 (.95) & 47.17 (3.92) & 26.67 (1.65)\\
        \bottomrule
    \end{tabular}
\end{table}
\begin{table}[!ht]
\captionsetup{skip=2pt}
\caption{Estimates for the Food Choice Model (Case 2a)}
\label{tab:fcm_2a}
\tiny
\setlength{\tabcolsep}{1.8pt}
   \centering
    \begin{tabular}{l l | c c | c c | c c}
        \multicolumn{8}{c}{\scriptsize{$(\Bar{\theta}_{SLT},\Bar{\theta}_{SUG},\Bar{\theta}_{SAT},\Bar{\theta}_{\text{variety}},\Bar{\theta}_{\text{skip}})= (.5,.5,.75,.1,9)$, $\beta = .90$}}\\
        \multicolumn{8}{c}{\scriptsize{no. of recipes ($M$) = 2, state-action space size = 30,001}}\\
        \toprule
        \multicolumn{2}{l|}{Parameter} & CCS & 1-step RLTS9-CCS & CCS & 1-step RLTD-CCS & CCS & 1-step RLTD-CCS \\
        \midrule
        \vspace{0.1cm}
        $T_{\text{end}}$ & & 5 & 5 & 10 & 10 & 50 & 50\\
        $\theta_{SLT}$ & Mean (Std.) & .5231 (2.72E-03) & .5011 (1.88E-04) & .4971 (2.43E-03) & .5000 (3.15E-05) & .4999 (4.55E-05) & .5000 (2.62E-05)\\
        \vspace{0.1cm}
         & RMSE & 2.32E-02 & 1.14E-03 & 3.76E-03 & 6.82E-05 & 4.54E-05 & 6.16E-05\\
        $\theta_{SUG}$ & Mean (Std.) & .469 (4.00E-03) & .5002 (8.21E-05) & .4757 (2.52E-03) & .5000 (2.87E-05) & .5000 (4.44E-05) & .5000 (3.14E-05)\\
        \vspace{0.1cm}
        & RMSE & .2015 & 7.07E-05 & 9.92E-02 & 6.29E-05 & 8.47E-05 & 7.25E-05\\
        $\theta_{SAT}$ & Mean (Std.) &  .7168 (3.62E-03) & .7504 (9.75E-05) & .7183 (2.86E-03) & .7500 (4.35E-05) & .7499 (6.14E-05) & .7500 (4.31E-05) \\
        \vspace{0.1cm}
        & RMSE & 3.33E-02 & 4.37E-04 & 3.17E-02 & 4.31E-05 & 6.37E-05 & 4.46E-05\\
        $\theta_{\text{variety}}$ & Mean (Std.) & -.0002 (2.75E-03) & .0998 (5.06E-05) & .065 (1.78E-03) & .0999 (3.41E-05) & .0999 (4.40E-05) & .0999 (2.89E-05)\\
        \vspace{0.1cm}
        & RMSE & .1003 & 1.42E-04 & 3.50E-02 & 3.48E-05 & 4.50E-05 & 2.97E-05\\
        $\theta_{\text{skip}}$ & Mean (Std.) & 6.076 (4.39E-02) & 9.0058 (1.25E-03) & 7.7147 (4.39E-02) & 9.0002 (5.27E-04) & 8.9997 (7.78E-04) & 9.0003 (5.66E-04)\\
        \vspace{0.1cm}
        & RMSE & 2.9214 & 5.98E-03 & 1.2859 & 5.84E-04 & 8.08E-04 & 6.38E-04\\
        \vspace{0.1cm}
        $\ell^2$-norm & Mean (Std.) & 4.6181 (5.69E-02) & 5.90E-02 (1.18E-02) & 2.8124 (2.81E-02) & 1.26E-02 (5.57E-03) & 4.46E-02 (5.37E-03) & 1.25E-02 (5.58E-03)\\
        \vspace{0.1cm}
        \#fevals & Mean (Std.) & 153.6 (4.99) & 329.52 (17.48) & 168 (3.83) & 364.58 (20.90) & 296.4 (8.39) & 370.04 (21.82)\\
        \vspace{0.1cm}
        Time (m) & Mean (Std.) & 1.64 (.11) & 2.33 (.26) & 3.17 (.09) & 3.99 (.24) & 24.29 (1.87) & 10.84 (1.16)\\
        \bottomrule
    \end{tabular}
\end{table}
\begin{table}[!ht]
\captionsetup{skip=2pt}
\caption{Estimates for the Food Choice Model (Case 2b)}
\label{tab:fcm_2b}
\tiny
\setlength{\tabcolsep}{1.8pt}
   \centering
    \begin{tabular}{l l | c c | c c | c c}
        \multicolumn{8}{c}{\scriptsize{$(\Bar{\theta}_{SLT},\Bar{\theta}_{SUG},\Bar{\theta}_{SAT},\Bar{\theta}_{\text{variety}},\Bar{\theta}_{\text{skip}}) = (.5,.5,.75,.1,9)$, $\beta = .90$}}\\
        \multicolumn{8}{c}{\scriptsize{no. of recipes ($M$) = 10, state-action space size = 550,011}}\\
        \toprule
        \multicolumn{2}{l|}{Parameter} & CCS & 1-step RLTD-CCS & CCS & 1-step RLTD-CCS & CCS & 1-step RLTD-CCS \\
        \midrule
        \vspace{0.1cm}
        $T_{\text{end}}$ & & 5 & 5 & 10 & 10 & 50 & 50\\
        $\theta_{SLT}$ & Mean (Std.) & .4126 (1.45E-03) & .5000 (4.86E-05) & .4325 (1.40E-03) & .5000 (4.59E-05) & .4999 (2.97E-05) & .5000 (4.03E-05)\\
        \vspace{0.1cm}
         & RMSE & 8.73E-02 & 5.52E-05 & 6.74E-02 & 4.79E-05 & 3.59E-05 & 4.59E-05\\
        $\theta_{SUG}$ & Mean (Std.) & .3233 (1.45E-03) & .4999 (3.69E-05) & .3981 (1.33E-03) & .4999 (3.63E-05) & .4999 (2.99E-05) & .4999 (2.87E-05)\\
        \vspace{0.1cm}
        & RMSE & .2015 & 7.07E-05 & 9.92E-02 & 6.29E-05 & 8.47E-05 & 3.10E-05\\
        $\theta_{SAT}$ & Mean (Std.) &  .6031 (1.61E-03) & .7500 (4.38E-05) & .6328 (1.42E-03) & .7500 (5.29E-05) & .7499 (2.99E-05) & .7500 (4.43E-05) \\
        \vspace{0.1cm}
        & RMSE & .1468 & 4.63E-05 & .1171 & 5.36E-05 & 5.56E-05 & 4.62E-05\\
        $\theta_{\text{variety}}$ & Mean (Std.) & .0213 (1.03E-03) & .1000 (2.44E-05) & .0541 (1.35E-03) & .1000 (3.87E-05) & .0999 (2.95E-05) & .1000 (3.06E-05)\\
        \vspace{0.1cm}
        & RMSE & 7.86E-02 & 2.64E-05 & 4.59E-02 & 3.97E-05 & 3.39E-05 & 3.12E-05\\
        $\theta_{\text{skip}}$ & Mean (Std.) & 3.516 (2.15E-02) & 9.0001 (6.59E-04) & 5.7506 (2.33E-02) & 9.0002 (8.50E-04) & 8.9991 (5.89E-04) & 9.0002 (6.84E-04)\\
        \vspace{.1cm}
        & RMSE & 5.4835 & 6.95E-04 & 3.2494 & 8.67E-04 & 1.06E-03 & 8.01E-04\\
        \vspace{0.1cm}
        $\ell^2$-norm & Mean (Std.) & 2.3888 (8.43E-03) & 5.60E-03 (2.58E-03) & 1.9362 (6.37E-03) & 3.82E-03 (1.58E-03) & 3.62E-02 (1.57E-04) & 3.71E-03 (1.56E-03)\\
        \vspace{0.1cm}
        \#fevals & Mean (Std.) & 125.76 (4.19) & 378.32 (14.54) & 136.32 (2.72) & 387.36 (14.95) & 240 (13.60) & 384.20 (19.54)\\
        \vspace{0.1cm}
        Time (m) & Mean (Std.) &  24.14 (3.98) & 52.77 (2.47) & 48.47 (3.81) & 84.26 (8.37) & 322.51 (18.83) & 191.77 (13.21)\\
        \bottomrule
    \end{tabular}
\end{table}
\begin{table}[!ht]
\captionsetup{skip=2pt}
\caption{Estimates for the Food Choice Model (Case 3a)}
\label{tab:fcm_3a}
\tiny
\setlength{\tabcolsep}{1.8pt}
   \centering
    \begin{tabular}{l l | c c | c c | c c}
        \multicolumn{8}{c}{\scriptsize{$(\Bar{\theta}_{SLT},\Bar{\theta}_{SUG},\Bar{\theta}_{SAT},\Bar{\theta}_{\text{variety}},\Bar{\theta}_{\text{skip}}) = (.5,.5,.75,.1,12)$, $\beta = .90$}}\\
        \multicolumn{8}{c}{\scriptsize{no. of recipes ($M$) = 2, state-action space size = 62,211}}\\
        \toprule
        \multicolumn{2}{l|}{Parameter} & CCS & 1-step RLTD-CCS & CCS & 1-step RLTD-CCS & CCS & 1-step RLTD-CCS \\
        \midrule
        \vspace{0.1cm}
        $T_{\text{end}}$ & & 5 & 5 & 10 & 10 & 50 & 50\\
        $\theta_{SLT}$ & Mean (Std.) & .5314 (2.26E-03) & .5023 (5.22E-04) & .5237 (1.19E-03) & .5000 (4.26E-05) & .4999 (2.95E-05) & .5000 (3.57E-05)\\
        \vspace{0.1cm}
         & RMSE & 3.15E-02 & 2.43E-03 & 2.38E-03 & 7.20E-05 & 2.95E-05 & 3.57E-05\\
        $\theta_{SUG}$ & Mean (Std.) & .4257 (2.66E-03) & .5007 (1.94E-04) & .4735 (2.19E-03) & .5000 (4.08E-05) & .4999 (3.63E-05) & .5000 (3.88E-05)\\
        \vspace{0.1cm}
        & RMSE & 7.43E-02 & 7.69E-04 & 2.65E-02 & 4.04E-05 & 4.82E-05 & 3.90E-05\\
        $\theta_{SAT}$ & Mean (Std.) &  .6714 (3.46E-03) & .7512 (3.14E-04) & .7246 (2.22E-03) & .7500 (5.31E-05) & .7499 (4.77E-05) & .7500 (5.10E-05) \\
        \vspace{0.1cm}
        & RMSE & 7.85E-02 & 1.27E-04 & 2.54E-02 & 5.30E-05 & 6.20E-05 & 5.26E-05\\
        $\theta_{\text{variety}}$ & Mean (Std.) & -.0454 (1.63E-03) & .0999 (6.15E-05) & .0539 (1.84E-03) & .0999 (2.57E-05) & .0999 (2.72E-05) & .1000 (2.34E-05)\\
        \vspace{0.1cm}
        & RMSE & .1454 & 1.19E-04 & 4.61E-02 & 2.78E-05 & 3.24E-05 & 2.37E-05\\
        $\theta_{\text{skip}}$ & Mean (Std.) & 9.1554 (5.57E-02) & 12.0252 (6.14E-03) & 11.5540 (3.99E-02) & 12.0004 (9.19E-04) & 11.9991 (7.42E-04) & 12.0005 (7.97E-04)\\
        \vspace{0.1cm}
        & RMSE & 2.8450 & 2.59E-02 & .4477 & 9.84E-04 & 1.13E-03 & 9.40E-04\\
        \vspace{0.1cm}
        $\ell^2$-norm & Mean (Std.) & 10.0032 (5.76E-02) & .1610 (3.58E-02) & 5.6062 (4.56E-02) & 3.59E-02 (2.91E-02) & 8.32E-02 (6.45E-04) & 3.58E-02 (2.91E-03)\\
        \vspace{0.1cm}
        \#fevals & Mean (Std.) & 163.92 (2.82) & 297.72 (11.61) & 183 (3.03) & 339.96 (13.85) & 314.16 (4.81) & 340.44 (13.96)\\
        \vspace{0.1cm}
        Time (m) & Mean (Std.) & 2.60 (.40) & 3.43 (.30) & 6.14 (.98) & 6.23 (1.02) & 47.73 (4.90) & 18.45 (2.23)\\
        \bottomrule
    \end{tabular}
\end{table}
\begin{table}[!ht]
\captionsetup{skip=2pt}
\caption{Estimates for the Food Choice Model (Case 3b)}
\label{tab:fcm_3b}
\tiny
\setlength{\tabcolsep}{1.8pt}
   \centering
    \begin{tabular}{l l | c c | c c | c c}
        \multicolumn{8}{c}{\scriptsize{$(\Bar{\theta}_{SLT},\Bar{\theta}_{SUG},\Bar{\theta}_{SAT},\Bar{\theta}_{\text{variety}},\Bar{\theta}_{\text{skip}}) = (.5,.5,.75,.1,12)$, $\beta = .90$}}\\
        \multicolumn{8}{c}{\scriptsize{no. of recipes ($M$) = 10, state-action space size = 1.14 million}}\\
        \toprule
        \multicolumn{2}{l|}{Parameter} & CCS & 1-step RLTD-CCS & CCS & 1-step RLTD-CCS & CCS & 1-step RLTD-CCS \\
        \midrule
        \vspace{0.1cm}
        $T_{\text{end}}$ & & 5 & 5 & 10 & 10 & 50 & 50\\
        $\theta_{SLT}$ & Mean (Std.) & .3844 (1.04E-03) & .5001 (4.35E-05) & .4376 (1.23E-03) & .5000 (3.78E-05) & .4999 (3.18E-05) & .5000 (5.15E-05)\\
        \vspace{0.1cm}
         & RMSE & .1155 & 1.30E-04 & 6.23E-02 & 4.83E-05 & 4.08E-05 & 7.67E-05 \\
        $\theta_{SUG}$ & Mean (Std.) & .2808 (1.04E-03) & .4999 (3.29E-04) & .3904 (1.32E-03) & .4999 (2.44E-05) & .4999 (2.74E-05) & .4999 (2.69E-05)\\
        \vspace{0.1cm}
        & RMSE & .2191 & 4.35E-05 & .1095 & 2.59E-05 & 5.06E-05 & 3.43E-05 \\
        $\theta_{SAT}$ & Mean (Std.) &  .5384 (1.77E-03) & .7500 (3.71E-05) & .6191 (1.52E-03) & .7500 (4.79E-05) & .7499 (2.78E-05) & .7499 (4.69E-05) \\
        \vspace{0.1cm}
        & RMSE & .2115 & 3.94E-05 & .1308 & 4.75E-05 & 6.45E-05 & 4.63E-05 \\
        $\theta_{\text{variety}}$ & Mean (Std.) & .0036 (9.19E-04) & .1000 (2.23E-05) & .0394 (8.81E-04) & .1000 (2.02E-05) & .0999 (2.03E-05) & .0999 (2.02E-05)\\
        \vspace{0.1cm}
        & RMSE & 9.63E-02 & 2.21E-05 & 6.05E-02 & 2.00E-05 & 2.92E-05 & 2.09E-05 \\
        $\theta_{\text{skip}}$ & Mean (Std.) & 5.0350 (2.90E-02) & 12.0003 (8.45E-04) & 8.2294 (3.32E-02) & 12.0000 (9.28E-04) & 11.9985 (7.88E-04) & 11.9998 (9.84E-05)\\
        \vspace{0.1cm}
        & RMSE & 6.9650 & 8.89E-04 & 3.7707 & 9.19E-04 & 1.67E-03 & 9.76E-04 \\
        \vspace{0.1cm}
        $\ell^2$-norm & Mean (Std.) & 4.5012 (1.30E-02) & 2.08E-02 (6.95E-03) & 3.6472 (1.12E-02) & 1.43E-02 (8.82E-03) & 6.94E-02 (1.094E-03) & 2.35E-02 (1.72e-02)\\
        \vspace{0.1cm}
        \#fevals & Mean (Std.) & 148.68 (2.51) & 256.3 (16.87) & 192.84 (2.42) & 365.86 (16.45) & 322.86 (12.65) & 355.09 (16.52)\\
        \vspace{0.1cm}
        Time (h) & Mean (Std.) & 1.03 (.17) & 1.79 (.15) & 1.98 (.17) & 2.25 (.27) & 13.28 (1.15) & 5.38 (.77)\\
        \bottomrule
    \end{tabular}
\end{table}
\begin{table}[!ht]
\captionsetup{skip=2pt}
\caption{Estimates for the Food Choice Model (Case 4a)}
\label{tab:fcm_4a}
\tiny
\setlength{\tabcolsep}{1.8pt}
   \centering
    \begin{tabular}{l l | c c | c c | c c}
        \multicolumn{8}{c}{\scriptsize{$(\Bar{\theta}_{SLT},\Bar{\theta}_{SUG},\Bar{\theta}_{SAT},\Bar{\theta}_{\text{variety}},\Bar{\theta}_{\text{skip}}) = (.5,.5,.75,.1,15)$, $\beta = .90$}}\\
        \multicolumn{8}{c}{\scriptsize{no. of recipes ($M$) = 2, state-action space size = 314,931}}\\
        \toprule
        \multicolumn{2}{l|}{Parameter} & CCS & 1-step RLTS9-CCS & CCS & 1-step RLTD-CCS & CCS & 1-step RLTD-CCS \\
        \midrule
        \vspace{0.1cm}
        $T_{\text{end}}$ & & 5 & 5 & 10 & 10 & 50 & 50\\
        $\theta_{SLT}$ & Mean (Std.) & .5136 (1.87E-03) & .5017 (2.79E-04) & .4615 (1.07E-03) & .5000 (2.92E-05) & .4999 (2.87E-05) & .5000 (3.30E-05)\\
        \vspace{0.1cm}
         & RMSE & 1.37E-02 & 1.80E-03 & 3.82E-02 & 8.02E-05 & 3.55E-05 & 8.57E-05\\
        $\theta_{SUG}$ & Mean (Std.) & .4625 (1.92E-03) & .5009 (1.58E-04) & .4187 (9.76E-04) & .5000 (4.85E-05) & .4999 (3.23E-05) & .5000 (2.68E-05)\\
        \vspace{0.1cm}
        & RMSE & 3.75E-02 & 9.46E-04 & 8.10E-02 & 5.31E-05 & 4.97E-05 & 4.01E-05\\
        $\theta_{SAT}$ & Mean (Std.) &  .6840 (2.73E-03) & .7513 (2.46E-04) & .6279 (1.54E-03) & .7500 (5.89E-05) & .7499 (4.31E-05) & .7500 (4.27E-05) \\
        \vspace{0.1cm}
        & RMSE & 6.60E-02 & 1.41E-03 & .1220 & 6.50E-05 & 7.40E-05 & 4.59E-05\\
        $\theta_{\text{variety}}$ & Mean (Std.) & .0073 (1.03E-03) & .1001 (2.49E-05) & .0699 (9.3E-03) & .1000 (1.91E-05) & .0999 (1.87E-05) & .1000 (1.35E-05)\\
        \vspace{0.1cm}
        & RMSE & 9.26E-02 & 1.11E-04 & 3.01E-02 & 1.90E-05 & 2.33E-05 & 1.42E-05\\
        $\theta_{\text{skip}}$ & Mean (Std.) & 16.9034 (9.01E-02) & 15.0360 (5.80E-03) & 14.4637 (4.27E-02) & 15.0011 (1.17E-03) & 14.9984 (9.60E-04) & 15.0012 (5.87E-04)\\
        \vspace{0.1cm}
        & RMSE & 1.9054 & 3.64E-02 & .5380 & 1.60E-03 & 1.88E-03 & 1.38E-03\\
        \vspace{0.1cm}
        $\ell^2$-norm & Mean (Std.) & 19.0901 (8.04E-02) & .2350 (2.96E-02) & 9.1915 (4.47E-02) & 6.47E-02 (3.03E-02) & .1590 (7.87E-04) & 6.46E-02 (3.04E-02)\\
        \vspace{0.1cm}
        \#fevals & Mean (Std.) & 257.54 (18.62) & 410.86 (20.36) & 314.56 (22.83) & 441.06 (27.75) & 419.68 (22.59) & 439.3 (26.15)\\
        \vspace{0.1cm}
        Time (m) & Mean (Std.) & 29.57 (3.76) & 33.81 (4.60) & 59.85 (7.11) & 50.56 (3.55) & 309.87 (26.32) & 112.86 (16.04)\\
        \bottomrule
    \end{tabular}
\end{table}
\begin{table}[!ht]
\captionsetup{skip=2pt}
\caption{Estimates for the Food Choice Model (Case 4b)}
\label{tab:fcm_4b}
\tiny
\setlength{\tabcolsep}{1.8pt}
   \centering
    \begin{tabular}{l l | c c | c c | c c}
        \multicolumn{8}{c}{\scriptsize{$(\Bar{\theta}_{SLT},\Bar{\theta}_{SUG},\Bar{\theta}_{SAT},\Bar{\theta}_{\text{variety}},\Bar{\theta}_{\text{skip}}) = (.5,.5,.75,.1,15)$, $\beta = .90$}}\\
        \multicolumn{8}{c}{\scriptsize{no. of recipes ($M$) = 10, state-action space size = 5.77 million}}\\
        \toprule
        \multicolumn{2}{l|}{Parameter} & CCS & 1-step RLTS9-CCS & CCS & 1-step RLTD-CCS & CCS & 1-step RLTD-CCS \\
        \midrule
        \vspace{0.1cm}
        $T_{\text{end}}$ & & 5 & 5 & 10 & 10 & 50 & 50\\
        $\theta_{SLT}$ & Mean (Std.) & .3437 (6.54E-04) & .5003 (4.04E-04) & .3753 (8.98E-04) & .5000 (2.83E-04) & * & *\\
        \vspace{0.1cm}
         & RMSE & .1562 & 5.48E-04 & .1246 & 2.79E-04 & * & *\\
        $\theta_{SUG}$ & Mean (Std.) & .3175 (6.5E-04) & .4994 (4.31E-04) & .3684 (6.26E-04) & .4996 (3.99E-04) & * & *\\
        \vspace{0.1cm}
        & RMSE & .1824 & 7.03E-04 & .1315 & 5.21E-04 & * & *\\
        $\theta_{SAT}$ & Mean (Std.) & .5178 (9.06E-04) & .7493 (7.68E-04) & .5603 (9.56E-04) & .7495 (6.91E-04) & * & * \\
        \vspace{0.1cm}
        & RMSE & .2321 & 9.85E-04 & .1896 & 8.11E-04 & * & *\\
        $\theta_{\text{variety}}$ & Mean (Std.) & .0108 (4.55E-04) & .0997 (2.31E-04) & .0493 (5.75E-04) & .0998 (2.69E-04) & * & *\\
        \vspace{0.1cm}
        & RMSE & 8.91E-02 & 3.72E-04 & 5.06E-02 & 3.26E-04 & * & *\\
        $\theta_{\text{skip}}$ & Mean (Std.) & 8.2640 (3.28E-02) & 14.9805 (2.05E-02) & 10.2325 (3.31E-02) & 14.9862 (1.96E-02) & * & *\\
        \vspace{0.1cm}
        & RMSE & 6.7360 & 2.81E-02 & 4.7676 & 2.33E-02 & * & *\\
        \vspace{0.1cm}
        $\ell^2$-norm & Mean (Std.) & 9.1118 (1.44E-02) & 3.29E-01 (1.03E-01) & 7.5770 (1.95E-02) & .2370 (.1070) & * & *\\
        \vspace{0.1cm}
        \#fevals & Mean (Std.) & 281.48 (12.26) & 363.1 (23.55) & 252.5 (12.90) & 381.5 (20.69) & * & *\\
        \vspace{0.1cm}
        Time (h) & Mean (Std.) & 7.60 (.44) & 7.59 (.59) & 13.63 (1.45) & 13.33 (1.00) & * & *\\
        \bottomrule
        \multicolumn{8}{l}{*estimation overran the maximum time limit of 24 hours on the high performance computing cluster}
    \end{tabular}
\end{table}

\subsection{Sensitivity to $\beta$}
We also found that RLTD-CCS was less sensitive to the discount factor, $\beta$. As $\beta$ increases, longer paths need to be simulated so that the impact of $\beta^{T_{\text{end}}}$ on the computed value functions becomes smaller than the simulation errors observed after averaging all the paths originating from a state. In RLTD-CCS, since the value functions are updated by looking one/$n$-steps ahead and as long as many iterative updates for all the states are performed, the sensitivity to $\beta$ is expected to be smaller than CCS. To confirm this, we estimated the smallest state-action space food choice model (3,891 states) for a $\beta$ of .995 and $T_{\text{end}}$ equal to 5, 10, and 500. 

The estimation results are presented in Table \ref{tab:fcm_beta}. The impact of a higher discount factor on the estimates from both the CCS and RLTD-CCS algorithms is clearly evident when compared to the results obtained with $\beta = .90$ (Table \ref{tab:fcm_1a}). The estimates for all the path lengths for both the algorithms were a little worse for the higher $\beta$ value. However, even at this higher discount factor, the 1-step RLTD-CCS algorithm achieved better RMSE performance for paths with $T_{\text{end}}$ equal to 10 than CCS achieves for paths that are 500 periods long. While the RMSE using CCS improves gradually as the path length increases, the RLTD-CCS algorithm achieves significant improvement when increasing the path length from 5 to 10, with only gradual gains beyond that point.

In addition, the memory requirements for storing paths of 500 periods were approximately 50 times greater than those for 10-period paths. These findings highlight that RLTD-CCS is less sensitive to $\beta$.

Overall, the results demonstrate that 1-step RLTD-CCS achieves comparable estimation accuracy while requiring shorter path lengths than CCS, resulting in significantly lower computation times even as the state-action space grows or the discount factor approaches one. These computational advantages suggest that RLTD-CCS can effectively extend the practical limits of ``forward simulation'' CCS estimation of DDC models to larger and more high-dimensional decision problems (e.g., incorporating more nutritional and non-nutritional attributes of recipes in the food choice model that could lead to more nuanced insights and granular counterfactual policy analysis).
\begin{table}[htp]
\captionsetup{skip=2pt}
\caption{Estimates for the Food Choice Model ($\beta=.995$)}
\label{tab:fcm_beta}
\tiny
\setlength{\tabcolsep}{1.8pt}
   \centering
    \begin{tabular}{l l | c c | c c | c c}
        \multicolumn{8}{c}{\scriptsize{$(\Bar{\theta}_{SLT},\Bar{\theta}_{SUG},\Bar{\theta}_{SAT},\Bar{\theta}_{\text{variety}},\Bar{\theta}_{\text{skip}}) = ( .5, .5, .75, .1, 5)$}}\\
        \multicolumn{8}{c}{\scriptsize{no. of recipes ($M$) = 2, state-action space size = 3,891}}\\
        \toprule
        \multicolumn{2}{l|}{Parameter} & CCS & 1-step RLTD-CCS & CCS & 1-step RLTD-CCS & CCS & 1-step RLTD-CCS \\
        \midrule
        \vspace{0.1cm}
        $T_{\text{end}}$ & & 5 & 5 & 10 & 10 & 500 & 500\\
        $\theta_{SLT}$ & Mean (Std.) &  .3765 (1.01E-02) &  .4174 (4.16E-03) &  .4014 (1.25E-02) &  .5004 (4.09E-04) &  .4999 (1.08E-03) &  .5000 (2.18E-04)\\
        \vspace{0.1cm}
         & RMSE & .1237 & 8.26E-02 & 9.93E-02 & 6.15E-04 & 1.03E-03 & 2.16E-04\\
        $\theta_{SUG}$ & Mean (Std.) &  .3928 (5.99E-03) &  .4226 (3.54E-03) &  .4057 (7.31E-03) &  .4993 (5.74E-04) &  .4991 (1.08E-03) &  .4998 (2.68E-04)\\
        \vspace{0.1cm}
        & RMSE & .1073 & 7.74E-02 & 9.44E-02 & 8.51E-04 & 1.24E-03 & 2.79E-04\\
        $\theta_{SAT}$ & Mean (Std.) &   .6082 (6.44E-03) &  .6434 (3.02E-05) &  .6226 (1.57E-02) &  .7492 (7.58E-04) &  .7495 (1.140E-03) &  .7497 (2.01E-04) \\
        \vspace{0.1cm}
        & RMSE & .1418 & .1065 & .1282 & 1.04E-03 & 1.42E-03 & 2.99E-04\\
        $\theta_{\text{variety}}$ & Mean (Std.) &  .0104 (6.84E-03) &  .0555 (3.94E-03) &  .0197 (9.49E-03) &  .0997 (3.2E-04) &  .0996 (8.19E-04) &  .1000 (1.87E-04)\\
        \vspace{0.1cm}
        & RMSE & 8.98E-02 & 4.46E-02 & 8.07E-02 & 3.93E-04 & 8.56E-04 & 1.79E-04\\
        $\theta_{\text{skip}}$ & Mean (Std.) &  2.5976 (6.98E-02) &  3.6382 (1.09E-02) &  3.2055 (.1087) &  4.9919 (6.92E-03) &  4.9951 (7.14E-03) &  .4998 (1.99E-03)\\
        \vspace{0.1cm}
        & RMSE & 2.4032 & 1.3618 & 1.7973 & 1.03E-02 & 8.33E-03 & 2.58E-03 \\
        \vspace{0.1cm}
        $\ell^2$-norm & Mean (Std.) & 1.3601 (2.80E-02) & .6697 (1.50E-02) & 1.4624 (3.60E-02) & 4.30E-02 (1.78E-02) & .1522 (2.65E-03) & 2.07E-02 (1.09E-02)\\
        \vspace{0.1cm}
        \#fevals & Mean (Std.) & 148.2 (2.89) & 222.3 (8.59) & 145.2 (3.79) & 257.7 (21.98) & 210.6 (9.96) & 289.2 (24.82)\\
        \vspace{0.1cm}
        Time (m) & Mean (Std.) & .1561 (5.12E-03) & .1647 (8.47E-03) & .3174 (1.62E-02) & .3011 (1.52E-02) & 23.068 (1.1582) & 9.32 (.80)\\
        \bottomrule
    \end{tabular}
\end{table}

\section{Conclusions}
As DDC models become even more high-dimensional, there is a pressing need for estimation methods that handle large state-action spaces efficiently. Simulations-based two-step CCS-like estimators, that are based on the CCP representation of the choice value functions, offer computational advantages over traditional full-solution approaches like NFXP. In this paper, we introduced a set of two-step forward simulation estimation algorithms, RLMC-CCS and RLTD-CCS that are inspired by the RL literature. These algorithms exploit visits to multiple state-action pairs during forward simulations to improve the efficiency of value function computation, addressing limitations of CCS. We showed that CCS is a special case of the RLMC-CCS algorithm and the CCS value function updates can be represented a geometric sums of RLTD-CCS updates. We provided Monte Carlo evidence to show the comparative advantage of the RL algorithms over CCS. Using a small state-space machine replacement model and large state-space food choice models (up to 5.77 million states), we showed that RLTD-CCS algorithms only need short simulation paths to reach the same level of estimation accuracy that was obtained from CCS on longer simulation paths. Using shorter paths for estimations helped speed up the computation by up to 14 times. We showed that the RL algorithms are less sensitive to discount factor than CCS. In one case, when when the discount factor was set to .995, RLTD-CCS was able to get accurate estimates 70 times faster than CCS.

Our work has implications for researchers in marketing, economics and computer science, as well as policy-makers, and practitioners. For researchers, our work provides computationally-light CCS-like estimation approaches for estimating large state-action space dynamic discrete-choice models. For practitioners, particularly who are working for eCommerce platforms where the state-action space within which consumers navigate is naturally large, we provide computationally faster methods to recovering consumer preferences and for carrying out counter-factual experiments. When the state-action space becomes extremely large (from millions to several billions), the step of pre-simulating and storing forward paths will quickly become impractical. In these scenarios, using a functional approximation strategy (both polynomial approximation and Deep Learning) might be better suited. Combining our proposed method with these functional approximation approaches may also be a worthwhile avenue to explore to seek further computational gains. Although, not the focus of our paper, in future it may be worthwhile to extend these algorithms to allow for unobserved heterogeneity. Another potential natural extension of our work is in the estimation of multi-agent DDC models, i.e., dynamic games.

In summary, we believe our extension of the CCS method using the RL algorithms is a promising step in overcoming computational challenges to estimating high-dimensional DDC models in a MDP framework, especially when the state-action space starts to become large. Our work draws on the synergy between the DDC and RL literatures. We hope this will help bridge the work in these two domains to help researchers combine their knowledge in advancing what are very similar research agendas.

\clearpage
\appendix
{\centering
\Large Online Appendix for Reinforcement Learning Based Computationally Efficient Conditional Choice Simulation Estimation of Dynamic Discrete Choice Models\par
\vspace{0.75em}
\normalsize Ahmed Khwaja \quad Sonal Srivastava\par
Judge Business School, University of Cambridge\par
\href{mailto:a.khwaja@jbs.cam.ac.uk}{a.khwaja@jbs.cam.ac.uk} \quad
\href{mailto:ss2450@cam.ac.uk}{ss2450@cam.ac.uk}\par
}

\section*{Appendix A: Value Function Computation Methods in DDC}\label{wa:valfunc}
In order to compute the CCPs for the state-action pairs, the value functions need to be computed first. There are two distinct classes of popular approaches to do this.

The first one is a \textit{full-solution} Nested Fixed Point Algorithm (``NFXP'') proposed by \cite{rust1987optimal}. His insight was to use the fixed point of the Bellman equation to compute the value functions. The choice-specific value function can be shown to be a contraction mapping $v = \Gamma(v)$ which, for the Type 1 MEV distribution (\textbf{Assumption~4}), can be expressed as:
\begin{equation}
    v(s,a; \theta) = u(s,a; \theta_u) + \beta \sum_{s^\prime} \mathit{p}(s^\prime | s, a; \theta_F) \log{\left(\sum_{a^\prime \in \mathcal{A}} e^{v(s^\prime,a^\prime; \theta)} \right)}
    \label{eq:contraction_mapping}
\end{equation}

In implementing NFXP, two further assumptions are required, Rational Expectations and stationarity (\cite{rust1994structural}, p.3083). Then the transition probabilities, $\hat{\mathit{p}}(s^\prime | s, a; \hat{\theta}_F)$, can be directly estimated from the observed data. Rational Expectations (\textbf{Assumption~5}) implies that given the information available to an agent their subjective beliefs about how the state variables evolve are consistent with the observed data. Further, in the infinite-horizon case, stationarity (\textbf{Assumption~6}) implies that the for the agent the ``future looks the same whether the agent is in state $s_t$ at time $t$ or in state $s_{t+k}$ at time $t+k$ provided that $s_{t+k} = s_t$'' (\cite{rust1994structural}, p.3091). Given functional form assumptions about payoffs $u(\cdot)$ and given the $i$\textsuperscript{th} guess, $\hat{\theta}^i_u$, the closed-form expression in Equation~\ref{eq:contraction_mapping} can be solved using the \textit{successive approximation} procedure (also called \textit{value iteration}) by starting with an initial guess of the value functions and then repeatedly sweeping through the entire state- space until a convergence criterion is met. When the discount-factor is sufficiently large ($\beta > .95$), the value iteration algorithm tends to slow down. \cite{rust1987optimal,rust1994structural} proposed a hybrid-approach (``polyalgorithm'') in the inner loop of his NFXP procedure to get around this issue. When the estimates of value functions are sufficiently close to the fixed-point, the inner loop switches to much faster Newton-Kantorovich iterations (equivalent to \textit{policy iteration}).

As the state space becomes large, this method of computing value function quickly becomes computationally demanding. To solve this issue, an \textit{approximate-solution} method was proposed by \cite{keane1994solution}. Their method uses polynomials to approximate the value function over a subset of the state-space while interpolating the value function on the remaining states, thereby, breaking the curse-of-dimensionality.

The second class of algorithms, which was first proposed by \cite{hotz1993conditional}, use an alternative representation of the choice-specific value functions (called the ``CCP'' representation). The insight behind this approach is that the CCPs have a one to one mapping to the normalized choice-specific value functions. Using Assumption 4, the normalized choice-specific value functions for a choice $a=m$ relative to, say, the first choice ($a=1$) can be written as:
\begin{equation}
    \Delta v(s,m; \theta) = v(s,m; \theta) - v(s,1; \theta) = \log{\left( \frac{\pi(m | s; \theta)}{\pi (1 | s; \theta)} \right)}
    \label{eq:one_to_one_map}
\end{equation}
One can non-parametrically estimate the CCPs ($\hat{\pi}(a | s)$) directly from the data and compute $\Delta \hat{v}(s,a)$. Asymptotically, $\Delta \hat{v}(s,a)$ is an unbiased estimate of $\Delta v(s,a; \theta)$. Building on this insight, \cite{hotz1994simulation}, proposed a simulation-based choice-specific value function estimator (``CCS''). Under this semi-parametric approach, the value function for a state-action pair can be computed by averaging returns over a number of forward simulated paths that originate from that state-action pair. The ``forward simulations'' themselves are done using the transition and choice probabilities non-parametrically estimated directly from the data. To elaborate, combing Equations 7 and 9 from the main text and then expanding the expectation term we can express the deterministic component of the choice-specific value function in a multi-period formulation as follows:
\begin{equation}
    v(s,a;\hat{\theta}^i_u,\hat{\theta}_F) = u(s,a;\hat{\theta}^i_u) + \beta \mathbb{E}\left[u(s^\prime,a^\prime; \hat{\theta}^i_u) + \epsilon(a^\prime) + \beta \mathbb{E} \left[ u(s^{\prime\prime}, a^{\prime\prime}; \hat{\theta}^i_u) + \cdots \right] \right]
    \label{eq:ccs_expand}
\end{equation}
The expectation of the error term can be written as \citep{hotz1993conditional}:
\begin{equation}
    \mathbb{E}\left[ \epsilon(a) | s, a\right] = \gamma - \log{(\hat{\pi}(a | s))},
    \label{eq:shock_spec}
\end{equation}
where, $\gamma \text{ }(=.57721...)$ is the Euler's constant. We can then evaluate the value function in Equation~\ref{eq:ccs_expand} by simulating $K$ forward paths using non-parametrically estimated transition and choice probabilities. For an infinite time-horizon case, the termination length for each path should be infinitely long. For practical purposes, however, the paths are terminated at a large $T_{\text{end}}$ beyond which, due to discounting, returns from the subsequent state-action pairs change the computed value function insignificantly. Once a $k$\textsuperscript{th} path has been simulated for a given starting state-action pair, the corresponding \textit{predicted} value function can be computed using the following expression:
\begin{multline}
    \tilde{v}^k(s,a;\hat{\theta}_u^i, \hat{\theta}_F) = u(s,a;\hat{\theta}_u^i) + \beta \bigg[u(s^\prime, a^\prime; \hat{\theta}_u^i) + \gamma - \log{(\hat{\pi}(a^\prime | s^\prime))} + \\
    \beta \Big[ u(s^{\prime \prime}, a^{\prime \prime}; \hat{\theta}_u^i) + \gamma - \log{(\hat{\pi}(a^{\prime\prime} | s^{\prime\prime}))} + \\
    \beta \big[...+\beta [u(s_{T_{\text{end}}},a_{T_{\text{end}}}; \hat{\theta}_u^i) + \gamma - \log{(\hat{\pi}(a_{T_{\text{end}}} | s_{T_{\text{end}}}))}]...\big]\Big]\bigg]
    \label{eq:ccs_formulation}
\end{multline}
The average value after computing the returns from $K$ paths can then be written as:
\begin{equation}
    \Bar{v}(s,a;\hat{\theta}_u^i, \hat{\theta}_F) = \frac{1}{K}\sum_{k=1}^K \tilde{v}^k(s,a; \hat{\theta}_u^i, \hat{\theta}_F)
    \label{eq:ccs_avg}
\end{equation}

This ``forward simulation'' based approach is advantageous over the full-solution approach as it skips the computationally-expensive fixed-point iterations. The main computational burden in the CCS approach arises from the forward simulations and the subsequent averaging operation. The CCP representation of value functions and the subsequent ``computationally lighter, two-step" CCS estimator has gained immense popularity and several extensions have been proposed in the literature \citep{aguirregabiria2002swapping,aguirregabiria2007sequential,bajari2007estimating,pesendorfer2008asymptotic}.
\enlargethispage{\baselineskip}

\section*{Appendix B: Estimation}\label{wa:estimation}
The aim of the researcher is to estimate the structural parameters corresponding to the utility function ($\theta_u$). However, that also requires estimating the parameters of the state transitions ($\theta_F$) from the observed data. In his NFXP algorithm, \cite{rust1987optimal} proposed a Maximum-Likelihood Estimator (MLE) with a \textit{log-likelihood} objective function that comprises two additively separable terms:
\begin{equation}
	LL(\theta) = \underbrace{\sum_{t=1}^{T} \log\left\lbrace
	\pi(a_{t} \big| s_{t}; \theta_u)\right\rbrace}_{\text{CCP term}} + \underbrace{\sum_{t=1}^{T} \log \left\lbrace \mathit{p}(s_{t} \big| s_{t-1}, a_{t-1}; {\theta_F}) \right\rbrace}_{\text{transition probability term}}
\end{equation}

Using Assumption 5 (Rational Expectations) and 6 (Stationarity), the transition probability term can be computed first. The CCP term is computed using nested loops. Under this routine, an outer loop uses a search algorithm to find the guess vector $\hat{\theta}^i_u$ that minimizes the log-likelihood function $LL(\hat{\theta}_u)$, and an inner loop that uses the polyalgorithm (value iterations followed by policy iterations) to compute, first, the value functions for a given guess vector (Equation~\ref{eq:contraction_mapping}) and then the corresponding CCPs. 

In the case of CCS, the estimation is done in two stages. In the first stage, non-parametric estimates for both the transition and choice probabilities are obtained from the data on observed states and choices, $\{(s_{\tau,n}^o,a_{\tau,n}^o)\}_{\tau=1,\ldots,T^o, n=1,\ldots,N}$, for $N$ agents. In the second stage, for a given guess $\hat{\theta}^i_u$, the predicted or ``simulated'' value functions, $\Bar{v}(s,a;\hat{\theta}^i_u,\hat{\theta}_F)$, are computed using Equation~\ref{eq:ccs_avg}. The utility parameters can be obtained by employing a Simulated Methods of Moments estimator using the normalized predicted and directly estimated value functions \citep{hotz1994simulation}:
\begin{align}
    \hat{\theta}^{i+1}_u &= \argmin_{\theta_u} \left(H_N(\hat{\theta}^i_u)^\prime \times W_N \times H_N(\hat{\theta}^i_u) \right),  \\
    \text{where, } \; \; H_N(\hat{\theta}^i_u) &= \frac{1}{N} \sum\limits_{n=1}^N \sum\limits_{t=1}^T \left[\Delta\Bar{v}(s_t^o,a_t^o;\hat{\theta}_u^i, \hat{\theta}_F) - \Delta\hat{v}(s_t^o,a_t^o)\right] Z_t^n .
\end{align}
Here $W_N$ is an $R \times R$ weighting matrix and $Z_t^n$ is an $R$ dimensional instrument vector.

A simpler, Minimum Distance Estimator (MDE), can also be used for estimating the parameters.  The MDE minimizes the distance between the predicted and directly estimated CCPs (or, value functions in an alternative version) \citep{bajari2007estimating}. The CCPs can be predicted using the dynamic logit formula after calculating the predicted value functions. Any standard optimizer routine can be used for generating the utility parameter guesses:
\begin{equation}
    \hat{\theta}_u^{i+1} = \argmin_{\theta_u} || \tilde{\pi}(a_t \vert s_t; \hat{\theta}_u^i, \hat{\theta}_F) - \hat{\pi}(a_t \vert s_t) ||
\end{equation}

The semi-parametric CCS approach has a computational advantage over NFXP.  A disadvantage of the CCS approach is that it is sensitive to the CCPs that are directly estimated from the data. Poor estimates of CCPs due to finite samples may lead to poor estimates of the model parameters. \cite{aguirregabiria2002swapping,aguirregabiria2007sequential} proposed the Nested Pseudo-Likelihood (NPL) estimator to improve the finite sample performance of such two-step estimators.

In problems involving large state-action spaces, NFXP can quickly become infeasible. While CCS (and other related methods) may be able to overcome this issue there has been limited demonstration of the CCS approach in dealing with such problems. An exception is \cite{bishop2012dynamic} who used a CCS based approach to estimate a large state space dynamic migration model relying on its ``finite dependence'' property \citep{arcidiacono2011conditional}. In contrast, our proposed approach doesn't rely on the finite dependence property.

\section*{Appendix C: RL Methods and DDC Modeling}\label{wa:rl}
RL algorithms are traditionally considered a separate class of machine learning methods when compared to supervised and unsupervised learning approaches. The aim of all machine learning methods is to help an agent (a computer program or a robot) learn from a training dataset and generate a response in situations that may or may not have been encountered before. Under supervised learning, an expert generates a labeled set of examples that are used to train the agent. The agent, when supplied with an input that was not present in the training dataset, assigns a label to the input depending on how similar it is to the training labels. While this approach has found tremendous success in audio-visual classifications, the learning itself is dependent on the richness of the training dataset. Under unsupervised learning, the agent is allowed to uncover hidden patterns in unlabeled data without any inputs from external experts. In both the cases, however, the learning is not motivated by the natural interaction of the agent with the environment in which it is embedded. One of the core tenets of RL is to allow the agent to take actions while it is interacting with its environment with the aim of fulfilling a well defined goal clearly specified by a reward function. This is done through a programmer encoding a reward for each action taken and the goal of the agent is to optimize its actions in order to maximize the total rewards. The dynamics of the agent-environment interaction is formalized as an MDP under which the agent transitions from one state to another by taking an action in every period.

While there are obvious similarities between the RL and DDC literatures (in both the cases, the underlying dynamics are assumed to be Markovian and both try to solve optimal control problems that were first introduced by \cite{bellman1952theory}), the aims are quite different. When researchers use RL methods (particularly in the fields of artificial intelligence and adaptive control), the aim is to find an optimal policy for the agent that could be used to navigate a fully/partially known and/or changing environment. On the other hand, in DDC, when modeling the decision-making behavior of an agent (typically, a consumer or manager of a firm), researchers assume that the agent is following an optimal policy (e.g., due to an utility or profit maximization assumption,  e.g., \cite{ellickson2012repositioning}). The aim instead is to solve the ``inverse'' problem of recovering the reward function of the agent (e.g., for counterfactual policy experiments).

An important step in solving an MDP is to calculate choice-specific value functions. The choice-specific value functions indicate how useful taking an action in a particular state will be. So, ``learning'' these is at the core of all the RL algorithms. A wide spectrum of RL algorithms have been developed to learn value functions. At one end of the spectrum, full-solution methods like RL Dynamic Programming (RL DP) exist. Classical RL DP methods (equivalent to NFXP), require a full model of the environment, i.e., assuming that the state transition probabilities are known to the researcher. At the other end of the spectrum are RL Monte Carlo (RL MC) methods which are ``model-free'' (in the RL jargon) and use real interactions between the agent and the environment to update the model (i.e., state transition probabilities) and learn the value functions. For a partially known environment model, RL uses planning methods like Dyna-Q \citep{sutton1991dyna}, that combine real interactions and forward simulations to compute value functions and obtain optimal policy(-ies). 

The RL DP algorithm computes the choice-specific value functions using a computationally expensive state-space sweep, similar to NXFP value function iterations. In RL jargon, the value function updates are called ``bootstrapped'' as in the value function at a given state is updated based on the estimated value functions of successor states. On the other hand RL MC methods compute the choice-specific value functions by taking averages of future-discounted returns over a large number of forward simulations/real interactions. In RL MC methods, the computation of choice-specific values function are not ``bootstrapped'' as these don't rely on the estimates of the value functions in successor states. Between extremes of RL DP and RL MC, exist a class of methods called Temporal-Difference (TD) learning \citep{sutton1988learning} that combine both forward simulations and ``bootstrapped'' value function updates. In fact, RL DP and RL MC methods can be treated as special cases of a general TD method.
\enlargethispage{\baselineskip}

\section*{Appendix D: Algorithms}\label{wa:algos}
All the algorithms (CCS, RLMC-CCS, RLTD-CCS) share a common, outer MDE loop that searches for the structural utility parameters, $\theta_u$ (Algorithm \ref{al:mde}). Below, we provide concise descriptions and pseudocode for the value function computation step in CCS, RLMC-CCS, and both 1-step and n-step variants of RLTD-CCS.
\begin{center}
\noindent\begin{minipage}{1\textwidth}
\begin{algorithm}[H]
\caption{Minimum Distance Estimator (MDE)}
\label{al:mde}
\scriptsize
\SetAlgoLined
	\textbf{Input:} $N_{\text{path}}$ forward simulated paths of length $T_{\text{end}}$, choice probabilities ($\hat{\pi}$), discount factor ($\beta$), and Euler's constant ($\gamma$)\\
    Initialize $\hat{\theta}^0_u$\\
    \While{$\hat{\theta}_u$ convergence criteria not met}{
    Compute $\tilde{v}(s,a; \hat{\theta}^i_u, \hat{\theta}_F)$ using CCS (Algorithm \ref{al:ccs})/RLMC-CCS (Algorithm \ref{al:rlmc-ccs})/RLTD-CCS (Algorithm \ref{al:1step_rltd-ccs}/\ref{al:nstep_rltd-ccs})\\
    Predict CCPs: $\tilde{\pi}(a = m \vert s;\hat{\theta}^i_u, \hat{\theta}_F) = \frac{e^{\tilde{v}(s,a=m; \hat{\theta}^i_u, \hat{\theta}_F)}}{\sum\limits_j e^{\tilde{v}(s,a=j; \hat{\theta}^i_u, \hat{\theta}_F)}} \; \; \forall \; \; a \in \mathcal{A}, s \in \mathcal{S}$\\
    Generate the next guess: $\hat{\theta}^{i+1}_u = \argmin\limits_{\theta_u} \vert\vert \tilde{\pi}(\cdot; \hat{\theta}^i_u, \hat{\theta}_F) - \hat{\pi}(\cdot) \vert\vert$
    }
\end{algorithm}
\end{minipage}
\end{center}

\subsection*{CCS}
Algorithm \ref{al:ccs} outlines the computation of the deterministic component of choice-specific value function, $\tilde{v}(s,a;\hat{\theta}_u^i, \hat{\theta}_F)$, in the ``Inner Loop'' of CCS for the $i$\textsuperscript{th} guess of structural utility parameters $\hat{\theta}_u^i$. The inputs to the algorithm include a set of $N_{\text{path}}$ simulated forward paths, each of length $T_{\text{end}}$; non-parametrically estimated CCPs $\hat{\pi}(\cdot)$ from the data; the discount factor $\beta$; and Euler's constant $\gamma$. These inputs are common across all the algorithms described in this section. 

The value function $\tilde{v}(s,a;\hat{\theta}_u^i, \hat{\theta}_F)$ is initialized to zero for all state-action pairs (line 2), and a visit counter $visits(s,a)$, which tracks the number of times a given pair $(s,a)$ appears as the starting pair in the simulated paths, is also initialized to zero (line 3). The algorithm then iterates over each simulated path (lines 4–12). For each path and its initial state-action pair, denoted $(s_1,a_1)$ in line 5, the visit count is incremented (line 6), and the cumulative utility term $G$ is initialized using the current utility $u(s_1,a_1;\hat{\theta}^i_u)$ (line 7). 
\begin{center}
\noindent\begin{minipage}{1\textwidth}
\begin{algorithm}[H]
\caption{CCS: Value Function Computation Step}
\label{al:ccs}
\scriptsize
\SetAlgoLined
	\textbf{Input:} $N_{\text{path}}$ forward simulated paths of length $T_{\text{end}}$, choice probabilities ($\hat{\pi}$), discount factor ($\beta$), and Euler's constant ($\gamma$) \\
    $\tilde{v}(s,a; \hat{\theta}^i_u, \hat{\theta}_F) = 0 \text{ } \; \; \forall \; \; s \in \mathcal{S}, a \in \mathcal{A}$\\
    $\text{visits}(s,a) = 0$\\
    \For{$\text{all }N_{\text{path}}$}{
    $(s_1,a_1) \leftarrow$ starting state-action pair\\
    $\text{visits}(s_1,a_1) \leftarrow \text{visits}(s_1,a_1) + 1$\\
    $G = u(s_1,a_1; \hat{\theta}^i_u)$\\
    \For{$t=2,3,...,T_{\text{end}}$}{
    $G  \leftarrow G  + \beta^{t-1}\left[u(s_t,a_t; \hat{\theta}^i_u)+\gamma-\log{\left(\hat{\pi}(a_t \vert s_t)\right)} \right]$\\
    }
    $\tilde{v}(s_1,a_1; \hat{\theta}^i_u, \hat{\theta}_F) \leftarrow \tilde{v}(s_1,a_1; \hat{\theta}^i_u, \hat{\theta}_F) + (G  - \tilde{v}(s_1,a_1; \hat{\theta}^i_u, \hat{\theta}_F))/\text{visits}(s_1,a_1)$
    }
    Return computed $\tilde{v}(s,a; \hat{\theta}^i_u, \hat{\theta}_F)$ to Algorithm \ref{al:mde} \\
\end{algorithm}
\end{minipage}
\end{center}

The algorithm then loops from $t=2$ to $T_{\text{end}}$ (line 8-10), and at each step, $G$ is updated by adding the both the deterministic and random components of the current utility (line 9). Once the full path is processed, the final value of $G$ is used to update the \textit{simulated} or \textit{predicted} $\tilde{v}(s_1,a_1;\hat{\theta}_u^i, \hat{\theta}_F)$ using the step-wise learning rule (line 11) as described in the main text. In this update, the (implicit) learning parameter $\alpha$ is given by the inverse of the visit count $visits(s_1,a_1)$. 

After all simulated paths have been processed, the algorithm returns the simulated value function $\tilde{v}(s,a; \hat{\theta}_u^i, \hat{\theta}_F)$, which is used in the outer MDE loop for generating the next guess $\hat{\theta}_u^{i+1}$.

\subsection*{RLMC-CCS}

Algorithm \ref{al:rlmc-ccs} describes the choice-specific value function computation step in RLMC-CCS, which extends CCS by updating the choice-specific value function not only for the starting state-action pair, but also for the subsequent pairs along a forward simulated path. As in CCS, we begin by computing the path-specific deterministic component of choice-specific value function which is equal to the final value of $G$ for a given path as outlined in lines 5-10 of the pseudocode. Once the final value of $G$ is calculated, $\tilde{v}(s_1,a_1;\hat{\theta}_u^i, \hat{\theta}_F)$ for the starting state-action pair $(s_1,a_1)$ is updated using the step-wise learning rule (line 11). These initial steps are identical to those used in the inner loop of the CCS algorithm, making CCS a special case of RLMC-CCS if the subsequent updates along the path in RLMC-CCS are omitted.

To update the value function for the subsequent state-action pairs along the path, RLMC-CCS uses a recursive strategy that exploits the algebraic structure of the discounted return. Let us consider the $k$\textsuperscript{th} forward simulated path. The path-specific deterministic component of the choice-specific value function for the starting state-action pair $(s_1,a_1)$, can be expressed as:
\begin{multline}
    \tilde{v}^{k,1}(s_1,a_1;\hat{\theta}^i_u, \hat{\theta}_F) \equiv G = u(s_1,a_1; \hat{\theta}^i_u) + \beta\left[u(s_2,a_2;\hat{\theta}^i_u) + \gamma - \log{\left(\hat{\pi}(a_2 \vert s_2)\right)}\right] + \\
    \beta^2\left[u(s_3,a_3;\hat{\theta}^i_u) + \gamma - \log{\left(\hat{\pi}(a_3 \vert s_3)\right)}\right] + \beta^3\left[u(s_4,a_4;\hat{\theta}^i_u) + \gamma - \log{\left(\hat{\pi}(a_4 \vert s_4)\right)}\right] + \cdots
    \label{eq:full_return}
\end{multline}

Similarly, the value function associated with the second state-action $(s_2, a_2)$ along the same path is given by:
\begin{multline}
    \tilde{v}^{k,2}(s_2,a_2;\hat{\theta}^i_u, \hat{\theta}_F) = u(s_2,a_2; \hat{\theta}^i_u) + \beta\left[u(s_3,a_3;\hat{\theta}^i_u) + \gamma - \log{\left(\hat{\pi}(a_3 \vert s_3)\right)}\right] + \\
    \beta^2\left[u(s_4,a_4;\hat{\theta}^i_u) + \gamma - \log{\left(\hat{\pi}(a_4 \vert s_4)\right)}\right] + \beta^3\left[u(s_5,a_5;\hat{\theta}^i_u) + \gamma - \log{\left(\hat{\pi}(a_5 \vert s_5)\right)}\right] + \cdots
\end{multline}

To express this in terms of the full-path return $G$, we multiply both sides of the above equation by $\beta$, and then add the term $\left(  u(s_1,a_1; \hat{\theta}^i_u) + \beta\left[\gamma - \log{\left(\hat{\pi}(a_2 \vert s_2)\right)}\right] \right)$ to both sides:
\begin{multline}
    u(s_1,a_1; \hat{\theta}^i_u) + \beta\left[\gamma - \log{\left(\hat{\pi}(a_2 \vert s_2)\right)}\right] + \beta \tilde{v}^{k,2}(s_2,a_2;\hat{\theta}^i_u, \hat{\theta}_F) = u(s_1,a_1; \hat{\theta}^i_u) + \\ \beta\left[u(s_2,a_2;\hat{\theta}^i_u) + \gamma - \log{\left(\hat{\pi}(a_2 \vert s_2)\right)}\right] + \beta^2\left[u(s_3,a_3;\hat{\theta}^i_u) + \gamma - \log{\left(\hat{\pi}(a_3 \vert s_3)\right)}\right] + \\ \beta^3\left[u(s_4,a_4;\hat{\theta}^i_u) + \gamma - \log{\left(\hat{\pi}(a_4 \vert s_4)\right)}\right] + \beta^4\left[u(s_5,a_5;\hat{\theta}^i_u) + \gamma - \log{\left(\hat{\pi}(a_5 \vert s_5)\right)}\right] + \cdots
\end{multline}

The right-hand side is equal to the full-path return $G$ as defined in Equation~\ref{eq:full_return}. Solving for $\tilde{v}^{k,2}(s_2,a_2;\hat{\theta}^i_u, \hat{\theta}_F)$ yields:
\begin{equation}
    \tilde{v}^{k,2}(s_2,a_2;\hat{\theta}^i_u, \hat{\theta}_F) = \frac{G - u(s_1,a_1; \hat{\theta}^i_u) - \beta\left[\gamma - \log{\left(\hat{\pi}(a_2 \vert s_2)\right)}\right]}{\beta}
\end{equation}
\begin{center}
\noindent\begin{minipage}{1\textwidth}
\begin{algorithm}[H]
\caption{RLMC-CCS: Value Function Computation Step}
    \label{al:rlmc-ccs}
\scriptsize
\SetAlgoLined
	\textbf{Input:} $N_{\text{path}}$ forward simulated paths of length $T_{\text{end}}$, choice probabilities ($\hat{\pi}$), discount factor ($\beta$), and Euler's constant ($\gamma$) \\
    $\tilde{v}(s,a; \hat{\theta}^i_u, \hat{\theta}_F) = 0 \text{ } \; \; \forall \; \; s \in \mathcal{S}, a \in \mathcal{A}$\\
    $\text{visits}(s,a) = 0$\\
    \For{$\text{all }N_{\text{path}}$}{
    $(s_1,a_1) \leftarrow$ starting state-action pair\\
    $\text{visits}(s_1,a_1) \leftarrow \text{visits}(s_1,a_1) + 1$\\
    $G = u(s_1,a_1; \hat{\theta}^i_u)$\\
    \For{$t=2,3,...,T_{\text{end}}$}{
    $G  \leftarrow G  + \beta^{t-1}\left[u(s_t,a_t; \hat{\theta}^i_u)+\gamma-\log{\left(\hat{\pi}(a_t \vert s_t)\right)} \right]$\\
    }
    $\tilde{v}(s_1,a_1; \hat{\theta}^i_u, \hat{\theta}_F) \leftarrow \tilde{v}(s_1,a_1; \hat{\theta}^i_u, \hat{\theta}_F) + (G  - \tilde{v}(s_1,a_1; \hat{\theta}^i_u, \hat{\theta}_F))/\text{visits}(s_1,a_1)$ \\
    \For{$t=2,3,...,T_{\text{end}}$}{
    $\text{visits}(s_t,a_t) \leftarrow \text{visits}(s_t,a_t) + 1$\\
    $G \leftarrow \left(G - u(s_{t-1},a_{t-1};\hat{\theta}^i_u)-\beta\left[\gamma - \log{\left(\hat{\pi}(a_t \vert s_t)\right)}\right] \right)/\beta$\\
    $\tilde{v}(s_t,a_t; \hat{\theta}^i_u, \hat{\theta}_F) \leftarrow \tilde{v}(s_t,a_t; \hat{\theta}^i_u, \hat{\theta}_F) + (G  - \tilde{v}(s_t,a_t; \hat{\theta}^i_u, \hat{\theta}_F))/\text{visits}(s_t,a_t)$ \\
    }
    }
    Return computed $\tilde{v}(s,a; \hat{\theta}^i_u, \hat{\theta}_F)$ to Algorithm \ref{al:mde} \\
\end{algorithm}
\end{minipage}
\end{center}

The sub-path-specific value function $\tilde{v}^{k,2}(s_2,a_2;\hat{\theta}^i_u, \hat{\theta}_F)$ can be interpreted as the full-path return value $G$ (line 14) corresponding to the sub-path starting at $(s_2,a_2)$. Once $G$ is updated in this way, the associated simulated value function $\tilde{v}(s_2,a_2;\hat{\theta}^i_u, \hat{\theta}_F)$ is updated using the same learning rule as before (line 15). These steps are repeated for each state-action pair along the simulated path, allowing $G$ and the associated value functions to be updated. 

Once all simulated paths have been processed and the simulated value function estimates $\tilde{v}(s,a;\hat{\theta}^i_u, \hat{\theta}_F)$ have been updated for each encountered state-action pair, the algorithm returns these values to the outer MDE loop.

\subsection*{RLTD-CCS}

In the RLTD-CCS algorithm, a fixed learning rate $\alpha$ is applied uniformly across all updates, eliminating the need to maintain a visit counter as in CCS or RLMC-CCS. The value functions are updated using ``bootstrapped'' temporal-difference learning in which the latest estimate of the value function for the successor state-action pair is used to update the value of the current state-action pair.

In the 1-step RLTD-CCS variant (Algorithm~\ref{al:1step_rltd-ccs}), the update at each time step uses the value function estimate of the immediate next state-action pair in the simulated path. Specifically, for each state-action pair $(s_t,a_t)$ in a path, the algorithm computes the 1-step temporal-difference (TD) error $\Delta_1(s_t,a_t; \hat{\theta}^i_u, \hat{\theta}_F)$ by comparing the current estimate $\tilde{v}(s_t,a_t; \hat{\theta}^i_u, \hat{\theta}_F)$ with a one-step ahead prediction. This includes the current utility $u(s_t,a_t; \hat{\theta}_u^i)$, discounted expectation of the next state-action pair's random utility component captured by $\gamma - \log{(\hat{\pi}(a_{t+1} \vert s_{t+1}}))$, and the discounted latest estimate of the value function for the next state-action pair, $\tilde{v}(s_{t+1},a_{t+1}; \hat{\theta}^i_u, \hat{\theta}_F)$ (line 5). The resulting TD error is then used to update $\tilde{v}(s_t,a_t; \hat{\theta}^i_u, \hat{\theta}_F)$ using a fixed learning parameter $\alpha$ (line 6).
\begin{center}
\noindent\begin{minipage}{1\textwidth}
\begin{algorithm}[H]
\caption{1-step RLTD-CCS: Value Function Computation Step}
    \label{al:1step_rltd-ccs}
\scriptsize
\SetAlgoLined
	\textbf{Input:} $N_{\text{path}}$ forward simulated paths of length $T_{\text{end}}$, choice probabilities ($\hat{\pi}$), discount factor ($\beta$), and Euler's constant ($\gamma$), learning parameter ($\alpha$)\\
    $\tilde{v}(s,a; \hat{\theta}^i_u, \hat{\theta}_F) = 0 \text{ } \; \; \forall \; \; s \in \mathcal{S}, a \in \mathcal{A}$\\
    \For{$\text{all }N_{\text{path}}$}{
    \For{$t=1,2,...,T_{\text{end}}-1$}{
    $\Delta_1(s_t,a_t; \hat{\theta}^i_u, \hat{\theta}_F) = u(s_t,a_t; \hat{\theta}^i_u) + \beta \left[\gamma - \log{\left(\hat{\pi}(a_{t+1} \vert s_{t+1})\right)}\right] + \beta\tilde{v}(s_{t+1},a_{t+1}; \hat{\theta}^i_u, \hat{\theta}_F)  - \tilde{v}(s_t,a_t; \hat{\theta}^i_u, \hat{\theta}_F)$ \\
    $\tilde{v}(s_t,a_t; \hat{\theta}^i_u, \hat{\theta}_F) \leftarrow \tilde{v}(s_t,a_t; \hat{\theta}^i_u, \hat{\theta}_F) + \alpha \Delta_1(s_t,a_t; \hat{\theta}^i_u, \hat{\theta}_F)$ \\
    }
    }
    Return computed $\tilde{v}(s,a; \hat{\theta}^i_u, \hat{\theta}_F)$ to Algorithm \ref{al:mde} \\
\end{algorithm}
\end{minipage}
\end{center}

The n-step RLTD-CCS algorithm (Algorithm~\ref{al:nstep_rltd-ccs}) generalizes this approach by incorporating longer look-ahead into the update. Instead of relying on only the immediate successor state-action pair, the update at step $t$ aggregates discounted returns over the next $n$ steps and uses the value function estimate at step $t+n$ as the ``bootstrap'' value. This results in an n-step TD error $\Delta_n(s_t,a_t; \hat{\theta}^i_u, \hat{\theta}_F)$ (line 5) which is then used for updating $\tilde{v}(s_t,a_t; \hat{\theta}^i_u, \hat{\theta}_F)$ using the fixed learning parameter $\alpha$. As in the 1-step case, updates are applied iteratively across each simulated path (lines 4-7). 

After all simulated paths have been processed, the updated value functions $\tilde{v}(s,a; \hat{\theta}^i_u, \hat{\theta}_F)$ are returned to the outer MDE loop to form the next guess of the structural utility parameters. 
\begin{center}
\noindent\begin{minipage}{1\textwidth}
\begin{algorithm}[H]
\caption{n-step RLTD-CCS: Value Function Computation Step}
    \label{al:nstep_rltd-ccs}
\scriptsize
\SetAlgoLined
	\textbf{Input:} $N_{\text{path}}$ forward simulated paths of length $T_{\text{end}}$, choice probabilities ($\hat{\pi}$), discount factor ($\beta$), and Euler's constant ($\gamma$), learning parameter ($\alpha$)\\
    $\tilde{v}(s,a; \hat{\theta}^i_u, \hat{\theta}_F) = 0 \text{ } \; \; \forall \; \; s \in \mathcal{S}, a \in \mathcal{A}$\\
    \For{$\text{all }N_{\text{path}}$}{
    \For{$t=1,2,...,T_{\text{end}}-n$}{
    $\Delta_n(s_t,a_t; \hat{\theta}^i_u, \hat{\theta}_F) = u(s_t,a_t; \hat{\theta}^i_u) + \beta \left[u(s_{t+1},a_{t+1}; \hat{\theta}^i_u) + \gamma - \log{\left(\hat{\pi}(a_{t+1} \vert s_{t+1})\right)}\right] + \cdots + \beta^n \tilde{v}(s_{t+n},a_{t+n}; \hat{\theta}^i_u, \hat{\theta}_F) - \tilde{v}(s_t,a_t; \hat{\theta}^i_u, \hat{\theta}_F)$ \\
    $\tilde{v}(s_t,a_t; \hat{\theta}^i_u, \hat{\theta}_F) \leftarrow \tilde{v}(s_t,a_t; \hat{\theta}^i_u, \hat{\theta}_F) + \alpha \Delta_n(s_t,a_t; \hat{\theta}^i_u, \hat{\theta}_F)$ \\
    }
    }
    Return computed $\tilde{v}(s,a; \hat{\theta}^i_u, \hat{\theta}_F)$ to Algorithm \ref{al:mde} \\
\end{algorithm}
\end{minipage}
\end{center}


\section*{Appendix E: Comparison of Value Function Update Frequency}\label{wa:updates}
Compared to CCS, the RLMC-CCS and RLTD-CCS algorithms make more efficient use of each simulated path by updating the value function not only for the starting state-action pair but also for pairs encountered along the path. As the number of simulated paths increases, the total number of value function updates grows accordingly. 

\begin{figure}[H]
    \centering
    \begin{subfigure}{0.4\textwidth}
        \centering
        \includegraphics[width=1\linewidth]{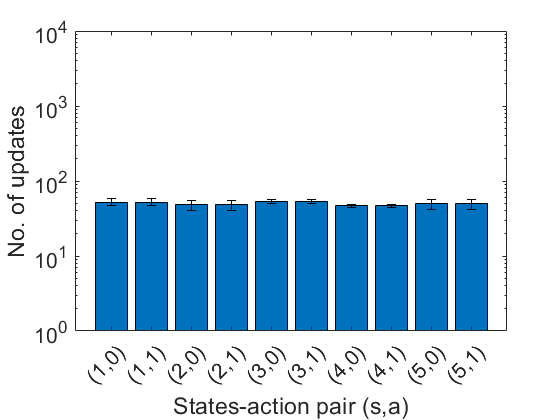}
        \caption{}        
    \end{subfigure}
    \begin{subfigure}{0.4\textwidth}
        \centering
        \includegraphics[width=1\linewidth]{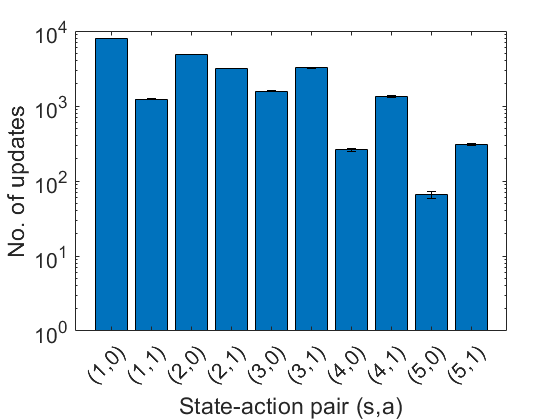}
        \caption{}
    \end{subfigure}
    \caption{\centering Average number of value function updates per state-action pair during a single execution of the inner loop for CCS and 1-step RLTD-CCS. The simulated path length was set to $T_{\text{end}}$ equal to 50, and a total of 500 paths were processed.}
    \label{fig:updates_compare}
\end{figure}

Figure~\ref{fig:updates_compare} shows the frequency with which the value function for each state-action pair is updated in the estimation of the Machine Replacement model, comparing CCS and 1-step RLTD-CCS. The comparison is based on a simulated path length of $T_{\text{end}}$ equal to 50 and a total of 500 paths. In a single execution of the inner loop, CCS updates the value function approximately 50 times per state-action pair. In contrast, 1-step RLTD-CCS updates the value function at every step along each path; the most frequently updated state-action pair, (1,0), receives nearly 8,000 updates, while even the least-visited pair, (5,0), is updated around 60 times.


\section*{Appendix F: Synthetic Data Generation for Food Choice Models}\label{wa:food}
Our aim was to evaluate the performance of the different algorithms for large state-action space sizes for the food choice model. The state-action space size of the food choice model is given by $(M \times H_{\text{max}} \times STOCK_{\text{max}}^3 + 1) \times (M + 1)$. We varied the state-action space size by changing the values of $STOCK_{\text{max}}$, $H_{\text{max}}$, and $M$. In total, we estimated eight different food choice models. The smallest model had a state-action space size of 3,891 and the largest model had a state-action space size of 5.77 million. The structural parameter values for each model were set to: $\Bar{\theta}_{SLT} = .5$, $\Bar{\theta}_{SUG} = .5$, $\Bar{\theta}_{SAT} = .75$, and $\Bar{\theta}_{\text{variety}} = .1$ while  $\Bar{\theta}_{\text{skip}}$ was set from 5 to 15 for different models. For the food choice models with a choice set of two recipes and a no order option, $r_m^{\text{fixed}}$ was set to .5 and .4 for $m = 1$ and 2 respectively. For the models with a larger choice set (\textit{i.e.}, ten recipes and a no order option), we randomly drew ten values for $r_m^{\text{fixed}}$ between .3 and .5. Similar to the machine replacement model, we used a discount factor value of .90. Using the assumed parameter values, we computed the choice-specific value functions for the state-action pairs using fixed-point iterations, followed by computing the corresponding CCPs. The computed CCPs and state transition dynamics were used to generate a unique synthetic dataset for each model. The panel length for each dataset was set to 100 days (unit of measurement in time was set to 1 day). To handle data sparsity issues in models, we varied the number of individuals from 100,000 (in the small state-action space size models) to 1 million (in the large state-action space size models). 

\bibliographystyle{agsm}
\bibliography{references.bib}

\end{document}